\documentclass[aps,pra,amssymb,twocolumn,amsmath,superscriptaddress,10pt,tightenlines]{revtex4-1}

\usepackage{graphicx}
\usepackage{dcolumn}
\usepackage{bm}

\usepackage[breaklinks, pdftex, hyperfootnotes=true, pdfpagelabels, bookmarks, pageanchor]{hyperref}

\hypersetup{%
	colorlinks=true, linktocpage=true, pdfstartpage=1, pdfstartview=FitH, pdfborder={0 0 0},%
	breaklinks=true, pdfpagemode=UseNone, pageanchor=true, pdfpagemode=UseOutlines,%
	plainpages=false, bookmarksnumbered, bookmarksopen=true, bookmarksopenlevel=1,%
	hypertexnames=true, pdfhighlight=/O,
	urlcolor=blue, linkcolor=blue, citecolor=blue, 
}

\usepackage{amssymb}
\usepackage{braket}
\usepackage{color}
\usepackage{here}
\usepackage{comment}
\usepackage{mathtools}
\mathtoolsset{showonlyrefs,showmanualtags}
\usepackage{overpic}

\begin{document}

\def\bra#1{\langle{#1}|}
\def\ket#1{|{#1}\rangle}
\def\sinc{\mathop{\text{sinc}}\nolimits}
\def\cV{\mathcal{V}}
\def\cH{\mathcal{H}}
\def\cT{\mathcal{T}}
\renewcommand{\Re}{\mathop{\text{Re}}\nolimits}
\newcommand{\tr}{\mathop{\text{Tr}}\nolimits}
\newcommand{\blambda}{\boldsymbol{\lambda}}

\definecolor{dgreen}{rgb}{0,0.5,0}
\newcommand{\green}{\color{dgreen}}
\newcommand{\BLUE}[1]{\text{\color{blue} #1}}
\newcommand{\GREEN}[1]{\textbf{\color{green}#1}}
\newcommand{\REV}[1]{\textbf{\color{red}[[#1]]}}
\newcommand{\KY}[1]{\textbf{\color{dgreen}[[#1]]}}

\def\HN#1{{\color{magenta}#1}}
\def\DEL#1{{\color{red}#1}}

\title{Generalized Adiabatic Impulse Approximation}

\author{Takayuki Suzuki}
\affiliation{Department of Physics, Waseda University, Tokyo 169-8555, Japan}
\author{Hiromichi Nakazato}
\affiliation{Department of Physics, Waseda University, Tokyo 169-8555, Japan}

\begin{abstract}
Non-adiabatic transitions in multilevel systems appear in various fields of physics, but it is not easy to analyze their dynamics in general. In this paper, we propose to extend the adiabatic impulse approximation to multilevel systems. This approximation method is shown to be equivalent to a series of unitary evolutions and facilitates to evaluate the dynamics numerically. In particular, we analyze the dynamics of the Landau--Zener grid model and the multilevel Landau--Zener--St\"uckelberg--Majorana interference model, and confirm that the results are in good agreement with the exact dynamics evaluated numerically. We also derive the conditions for destructive interference to occur in the multilevel system.
\end{abstract}

\maketitle

\section{Introduction\label{sec:intro}}
The adiabatic approximation is a well-known approximation technique in quantum mechanics. According to the approximation, the state at any time can be regarded as the $n$th instantaneous eigenstate of the slowly varying time-dependent Hamiltonian if the initial state has been prepared in the $n$th eigenstate of the initial Hamiltonian~\cite{Born1928-lj,Kato1950-me,Messiah2014-bx}. This adiabatic approximation has been used in various fields of physics: quantum adiabatic computation~\cite{Albash2016-wl,Barends2016-ct} and quantum control~\cite{Kral2007-jz}. If the Hamiltonian changes but not slowly enough, the adiabatic approximation breaks down and transitions between instantaneous eigenstates occur. Such transitions are called non-adiabatic transitions, and the Landau--Zener (LZ) model is known as the simplest and useful model of the non-adiabatic transition in the time-dependent two-level system.  Its Hamiltonian has diagonal elements that depend linearly on time, while off-diagonal elements are time-independent~\cite{Landau1932-uy,Zener1932-ve,Stuckelberg1932-jk,Majorana1932-xs}. The dynamics of such non-adiabatic transitions has attracted much attention in various fields and physical systems: quantum information science~\cite{quintana2013cavity,matityahu2019dynamical}, chemical physics~\cite{Nitzan2020-dn},  atomic physics~\cite{salger2007atomic,troiani2017landau,zhang2018symmetry,Niranjan2020-ta}, circuit-QED system~\cite{Chiorescu2004-st,Wallraff2004-op}, ultracold atom system~\cite{Kohler2006-lr}, and quantum dot system~\cite{petta2010coherent}. In general, however, solving the Schr\"odinger equation for the time-dependent multilevel system is difficult. Therefore, various approximations, including adiabatic approximation, have been adopted to analyze the dynamics. In particular, the adiabatic impulse approximation (AIA) is known as a method that enables us to treat the dynamics analytically~\cite{Kayanuma1997-ec,Shevchenko2010-ic}. This is an approximation under which the system evolves adiabatically except when the energy gap becomes small and non-adiabatic transition occurs instantaneously. Since it requires information on instantaneous eigenvalues, this approximation can only be used for relatively small systems such as two- and three-level systems~\cite{Niranjan2020-ta,Ostrovsky2007-dg}.

A general method to analytically approximate the dynamics of a multilevel system has not been known so far. On the other hand, there are physical systems with many levels in which non-adiabatic transitions between them play important roles. One of them is known as the multilevel LZ model, in which the diagonal elements of the Hamiltonian depend linearly on time, and the off-diagonal elements are independent of time. Another important example is the multilevel Landau--Zener--St\"uckelberg--Majorana (LZSM) interference model, in which the diagonal elements of the Hamiltonian are periodic functions of time, while the off-diagonal elements are constants. These models are realizations of such physical systems as Rydberg atom systems~\cite{forre2003selective,harmin1994incoherent,harmin1997coherent}, circuit QED systems~\cite{werther2019davydov,sun2012photon,huang2018dynamics,malla2018landau,Keeling2008-jo,Lidal2020-go,Wang2021-fy,Zheng2021-qb,Satanin2012-jf,Du2010-sf,Neilinger2016-jw,Gramajo2019-xy,Bonifacio2020-lr,Parafilo2018-gc}, quantum dot systems~\cite{Forster2014-ht,Shevchenko2018-ay,Mi2018-rf,petta2010coherent,Danon2014-lk,Ribeiro2013-ag,Stehlik2014-qr,Stehlik2016-py,Pasek2018-nb}, and many-body spin systems~\cite{Sinitsyn2013-qo,Wang2008-rd,Ostrovsky2006-zk}. It is thus of essential importance to have appropriate methods to analyze these models.

In this paper, we propose a generalization of AIA (generalized AIA, GAIA) where the scope of AIA is extended to general multilevel systems. For this purpose, we use the idea of exact WKB analysis of previous studies~\cite{Aoki2002-zw,shimada2020numerical}. We will show that the idea leads to a succession of local unitary transitions and see that the result is an extension of the conventional AIA. We also illustrate that GAIA can be used for situations where the applicability conditions of previous studies are not necessarily met.

The structure of this paper is as follows. In Sec.~\ref{sec:review}, we review the derivation of the S-matrix within the framework of exact-WKB analysis. In Sec.~\ref{sec:LZ_grid}, we extend the idea of AIA to the LZ grid model (GAIA) and show that the derived S-matrix agrees well with numerical calculations within a valid parameter region of approximation. In Sec.~\ref{sec:LZSM}, we derive the S-matrix for the multilevel LZSM interference model by referring to the correspondence between the GAIA derived in Sec.~\ref{sec:LZ_grid} and previous studies. We show that the results agree with numerical calculations within a valid parameter region of approximation. We also derive the conditions for destructive interference. Finally, a short summary is given in Sec.~\ref{sec:conclusion}.

\section{review : Exact WKB analysis for three-level LZ model\label{sec:review}}
In this section, we present a brief review of the exact-WKB analysis for the three-level LZ model in previous studies~\cite{Aoki2002-zw,shimada2020numerical}. Consider the following Schr\"odinger equation ($\hbar=1$):
\begin{align}
    i \frac{\partial }{\partial t}\ket{\psi(t,\eta)}=& H(t,\eta) \ket{\psi(t,\eta)},\label{eq:prev_sch}\\
    H(t,\eta)=&\eta
    \left(\begin{array}{ccc}
    \rho_1(t) & 0 & 0 \\
    0 & \rho_2(t) & 0 \\
    0 & 0 & \rho_3(t)
    \end{array}\right)\\
    &+\eta^{1 / 2}
    \left(\begin{array}{ccc}
    0 & b_{12} & b_{13} \\
    b_{12}^\ast & 0 & b_{23} \\
    b_{13}^\ast & b_{23}^\ast & 0
    \end{array}\right)\\
    =:&\eta (H_0(t)+\eta^{-1/2}H_{1/2}(t))\label{eq:prev_ham}.
\end{align} 
Here $\eta$ (essentially equal to $1/\hbar$) is a large parameter (adiabatic parameter). 
We endeavor to find the S-matrix whose elements $S_{i,j}$ give the transition amplitudes up to phase between basis states $\ket i$ and $\ket j$ from $t=-\infty$ to $t=\infty$: $|S_{i,j}|=|\braket{i|U(\infty,-\infty)|j}|$, where $U$ is the time-evolution operator generated by the Hamiltonian~\eqref{eq:prev_ham}. For this purpose, we construct a global WKB solution and local WKB solutions around the anti-crossing points and connect them. 

To find the global WKB solution, we formally diagonalize the Hamiltonian~\eqref{eq:prev_ham}: a unitarily transformed state $|\varphi(t,\eta)\rangle$ from $|\psi(t,\eta)\rangle$, $\ket{\psi(t,\eta)}=\left(1+\eta^{-1 / 2} P_{1 / 2}(t)\right)\left(1+\eta^{-1} P_{1}(t)\right) \cdots\ket{\varphi(t,\eta)}$, satisfies
\begin{widetext}
\begin{align}
    i \frac{\partial}{\partial t} \ket{\varphi(t,\eta)}=&\eta\left(H_{0}(t)+\eta^{-1} \tilde{H}_{1}(t)+\eta^{-3 / 2} \tilde{H}_{3 / 2}(t)+\cdots\right) \ket{\varphi(t,\eta)},
\end{align}
where the matrices $P_i(t)$ are defined  recursively~\cite{Aoki2002-zw} to make each $\tilde H_i(t)$ diagonal.
Then the formal solution of \eqref{eq:prev_sch} that does not suffer from transitions to other states from state $|j\rangle$, the so-called global WKB solution, reads as
\begin{align}
    \ket{\psi^{(j)}(t,\eta)}=\exp \left(-i\eta \int^{t} \rho_{j}(t) d t-i \int^{t}\sum_{k\not= j}\frac{\left|b_{j k}\right|^{2}}{\rho_{j}(t)-\rho_{k}(t)}dt\right)\ket{j}+O(\eta^{-1}).
\end{align}
To calculate the S-matrix, we need to introduce  normalization phase-factor (diagonal) matrices $N^{(\pm)}$ of the global WKB solution at $t\to\pm\infty$:
\begin{align}
    \left(\ket{j},\ket{k},\ket{l}\right)=&  \left(\ket{\psi^{(j)}(-\infty,\eta)},\ket{\psi^{(k)}(-\infty,\eta)},\ket{\psi^{(l)}(-\infty,\eta)}\right)N^{(-)},\\
    \left(\ket{j},\ket{k},\ket{l}\right)=&  \left(\ket{\psi^{(j)}(\infty,\eta)},\ket{\psi^{(k)}(\infty,\eta)},\ket{\psi^{(l)}(\infty,\eta)}\right)N^{(+)}.
\end{align}
\end{widetext}

Next, we proceed to find the local WKB solution.  Since the Schr\"odinger equation for the two-level LZ model reduces to the Weber equation, we naively assume that the multilevel system can also be described by the Weber equation in the vicinity of the anti-crossing point, but in reality, a more detailed discussion is needed. For this purpose, we first introduce the notion of turning point and Stokes line: the time $t=\tau$ at which $\rho_j(\tau)=\rho_k(\tau)$ is satisfied is defined as the turning point of type $(j,k)$. The corresponding Stokes curve is defined as such (complex) $t$ that satisfies the condition
\begin{align}
    \operatorname{Im}\left(-i \int_{\tau}^{t}\left(\rho_{j}(t')-\rho_{k}(t')\right) d t'\right)=0.
\end{align}
It is known that the behavior of the asymptotic series changes when the analytic continuation is made crossing over the Stokes curves, and the behavior of the asymptotic series for second-order differential equations such as the Weber function has been well studied. The matrix that relates behavior of the asymptotic series before and after crossing the Stokes curve is called the connection matrix.

In the three-level LZ model, when we consider a vicinity of a turning point, we can consider the Stokes curve in the vicinity. On the other hand, when discussing the global behavior, it is generally necessary to consider other Stokes curves called ``new Stokes curves"~\cite{symposium1994analyse,aoki1998exact,honda2015virtual}. It is, however, known that the new Stokes curves do not contribute to the S-matrix if the following reality condition is satisfied~\cite{Aoki2002-zw,shimada2020numerical}:
\begin{widetext}
\begin{align}
\left(\rho_{1}(t)-\rho_{2}(t)\right)\left(\rho_{2}(t)-\rho_{3}(t)\right)\left(\rho_{3}(t)-\rho_{1}(t)\right)=0\text{ has only real and simple zeros}.
\end{align}
In this paper, we consider the cases which do not violate the reality condition or the cases where we need to consider only the ordinary Stokes curves even if the reality condition is violated. Therefore, we need to consider the connection matrix when crossing the Stokes curve near the turning point on the real axis. We define $t_{j,k}^{[n]}$ as the $n$th turning point of type $(j,k)$ and assume 
\begin{align}
    \lambda_{j,k}^{[n]}:=\left.\frac{d}{dt}(\rho_k(t)-\rho_j(t))\right|_{t=t_{j,k}^{[n]}}>0.
\end{align}
The connection matrix $M_{jk}$ responsible for crossing Stokes curve near the turning point at $t=t_{j,k}^{[n]}$ can be expressed as
\begin{align}
    \left(\ket{\psi^{(j)}(t,\eta)},\ket{\psi^{(k)}(t,\eta)},\ket{\psi^{(l)}(t,\eta)}\right)\mapsto& \left(\ket{\psi^{(j)}(t,\eta)},\ket{\psi^{(k)}(t,\eta)},\ket{\psi^{(l)}(t,\eta)}\right)\begin{pmatrix}
    p_{jk}^{[n]}&-\alpha_{jk}^{[n],+}&0\\
    -\alpha_{jk}^{[n],-}&1&0\\
    0&0&1
    \end{pmatrix}\\
    =&: \left(\ket{\psi^{(j)}(t,\eta)},\ket{\psi^{(k)}(t,\eta)},\ket{\psi^{(l)}(t,\eta)}\right)M_{jk},
\end{align}
where
\begin{align}
\alpha_{j k}^{[n], \pm}=&(2 \eta)^{\kappa_{j k}^{[n]}} \sqrt{\frac{2 \pi}{\lambda_{j k}^{[n]}}} \frac{b_{j k}^{\pm}}{\Gamma\left(1+\kappa_{j k}^{[n]}\right)} e^{i \pi(1 / 2 \mp 1)\left(\kappa_{j k}^{[n]} \mp 1 / 2\right)}  \left(\beta_{j k}^{[n]}\right)^{\pm 1}\left(1+O\left(\eta^{-1 / 2}\right)\right),\\
b_{j k}^{+}=&b_{j k}, \quad b_{j k}^{-}=b^\ast_{j k}, \quad \kappa_{j k}^{[n]}=\frac{i\left|b_{j k}\right|^{2}}{\lambda_{j, k}^{[n]}},\\
\beta_{j k}^{[n]}=&e^{-i\eta \int_{t_{ j k}^{[n]}}^{t_{0}}\left(\rho_{j}\left(t^{\prime}\right)-\rho_{k}\left(t^{\prime}\right)\right) d t^{\prime}} \prod_{\substack{m=1\\m\neq n}}^{p}\left(t_{j k}^{[n]}-t_{j k}^{[m]}\right)^{-2 \kappa_{jk}^{[m]}} \prod_{\substack{m=1\\m\neq n}}^{q}\left(t_{j k}^{[n]}-t_{j l}^{[m]}\right)^{-\kappa_{j l}^{[m]}}  \prod_{\substack{m=1\\m\neq n}}^{r}\left(t_{j k}^{[n]}-t_{k l}^{[m]}\right)^{\kappa_{k l}^{[m]}}\left(\frac{\lambda_{j k}^{[n]}}{2}\right)^{\kappa_{j k}^{[n]}},
\end{align}
with $p,q$, and $r$ expressing  the numbers of times that  $\rho_j(t)$ and $\rho_k(t)$,  $\rho_j(t)$ and $\rho_l(t)$, and $\rho_k(t)$ and $\rho_l(t)$ cross, respectively.
In the case of $\rho_1(t)=v_1t+a,\rho_2(t)=v_2t,\rho_3(t)=v_3t$ where $a>0$ and $b_3>b_2>b_1>0$, the connection matrix is given by~\cite{Aoki2002-zw}
\begin{align}
    \left(\ket{\psi^{(j)}(t,\eta)},\ket{\psi^{(k)}(t,\eta)},\ket{\psi^{(l)}(t,\eta)}\right)\mapsto&  \left(\ket{\psi^{(j)}(t,\eta)},\ket{\psi^{(k)}(t,\eta)},\ket{\psi^{(l)}(t,\eta)}\right)M_{12}M_{13}M_{23}.
\end{align}
The S-matrix (under AIA) can be represented as
\begin{align}
    S=&(N^{(+)})^{-1}M_{12}M_{13}M_{23}N^{(-)}\\
    =&\left(\begin{array}{ccc}
    e^{i \pi\left(\kappa_{12}+\kappa_{13}\right)} & \alpha_{23}^{-} \alpha_{13}^{+} e^{i \pi\left(2 \kappa_{12}-\kappa_{23}\right)}-\alpha_{12}^{+} e^{i \pi \kappa_{23}} & -\alpha_{13}^{+} e^{2 i \pi \kappa_{12}}+\alpha_{12}^{+} \alpha_{23}^{+} \\
    -\alpha_{12}^{-} e^{i \pi \kappa_{13}} & -\alpha_{12}^{-} \alpha_{23}^{-} \alpha_{13}^{+} e^{i \pi\left(\kappa_{12}-\kappa_{23}\right)}+e^{i \pi\left(\kappa_{12}+\kappa_{23}\right)} & \left(\alpha_{12}^{-} \alpha_{13}^{+}-\alpha_{23}^{+}\right) e^{i \pi \kappa_{12}} \\
    -\alpha_{13}^{-} e^{i \pi\left(-\kappa_{12}+\kappa_{23}\right)} & -\alpha_{23}^{-} e^{i \pi \kappa_{13}} & e^{i \pi\left(\kappa_{23}+\kappa_{13}\right)}
    \end{array}\right).
\end{align}
\end{widetext}

\section{Landau--Zener grid model\label{sec:LZ_grid}}
\subsection{GAIA\label{subsec:gaia}}
This section considers the LZ grid model, which is a particular case of multilevel LZ models and has two parallel energy bands that cross over time in the energy diagram. We remark that the discussion in this section is also applicable to the general multilevel LZ model. Consider a $2N$-level system governed by the Hamiltonian 
\begin{align*}
    H(t)=&\eta A+\sqrt{\eta}B,\\
    A=&\operatorname{diag}(-vt+a_1,\cdots,-vt+a_N\\
    &\quad\quad\quad\quad,vt+a_1,\cdots,vt+a_N),\\
    B=&\begin{pmatrix}
    0&\cdots&0&b_{1,N+1}&\cdots&b_{1,2N}\\
    \vdots&\ddots&\vdots&\vdots&\ddots&\vdots\\
    0&\cdots&0&b_{N,N+1}&\cdots&b_{N,2N}\\
    b_{1,N+1}^\ast&\cdots&b_{N,N+1}^\ast&0&\cdots&0\\
    \vdots&\ddots&\vdots&\vdots&\ddots&\vdots\\
    b_{1,2N}^\ast&\cdots&b_{N,2N}^\ast&0&\cdots&0
    \end{pmatrix},
\end{align*}
where $\eta$ is a large parameter~\cite{Aoki2002-zw,shimada2020numerical}, the physical meaning of which will be explained later. $v$ is a ramp parameter, $a_k$ are responsible for level spacings, and $b_{ij}$ are couplings between $i$th and $j$th levels.

The LZ grid model has applications in various fields, including atomic physics~\cite{harmin1994incoherent,harmin1997coherent}, quantum information science~\cite{sun2012photon,malla2017loss}, and open quantum physics~\cite{sinitsyn2003nuclear,garanin2008effect,Wubs2006-do,Saito2007-fv,ashhab2014landau}. No general method, however, is known to analyze the transition probabilities in the LZ grid model. Approximate methods have been developed for the case where the separation of parallel levels in a band is very small \cite{yurovsky2001curve}. Besides, there are several arguments against the validity of transition probabilities in the LZ grid model~\cite{Usuki1997-gz,Wilkinson2000-fd,malla2017loss,malla2021nonadiabatic}.

The S-matrix in this model can be obtained approximately as the following matrix $S$ in the same way as in the discussion in \cite{Aoki2002-zw,shimada2020numerical}:
\begin{align}
    S=&\left(\tilde N^{(+)}\right)^{-1} \tilde M_{1}\cdots\tilde M_{2N-2}\tilde M_{2N-1} \tilde N^{(-)},\label{eq:prev_S}\\
    \tilde M_{k}=&\prod_{\substack{j-i=k \\ 1\leq i\leq N, N+1\leq j\leq 2N}}M_{i,j},\\
    M_{i, j}=&\left(\begin{array}{ccccc}
    I_{i-1} & & & & \\
     & p_{i ,j} & \cdots & -\alpha_{i ,j}^{+} & \\
     & \vdots & I_{j-i-1} & \vdots & \\
     & -\alpha_{i ,j}^{-} & \cdots & 1 & \\
     & & & & I_{2 N-j}
    \end{array}\right),\\
    \tilde N^{(+)}=&\prod_{k=1}^{N}\left(\begin{array}{ccccc}
    I_{N} & & & & \\
     &  p_{k, N+1}^{-1 / 2}&  &  & \\
     &  &  p_{k, N+2}^{-1 / 2} &  & \\
     &  & & \ddots & \\
     & & & &  p_{k, 2N}^{-1 / 2}
    \end{array}\right),\label{eq:n_plus}\\
    \tilde N^{(-)}=&\prod_{k=N+1}^{2N}\left(\begin{array}{ccccc}
    p_{1, k}^{-1 / 2} & & & & \\
     &  p_{2, k}^{-1 / 2}&  &  & \\
     &  &  \ddots &  & \\
     &  & & p_{N, k}^{-1 / 2} & \\
     & & & &  I_N
    \end{array}\right),\\
    p_{i, j}=&e^{-2 \pi \kappa_{i ,j}},\quad  \kappa_{i, j}=\frac{\left|b_{i, j}\right|^{2}}{2 v},\\
    \alpha_{i ,j}^{\pm}
    =&\pm \sqrt{1-p_{i,j}}\\
    &\times e^{\pm i\left( \pi / 4+ \arg \Gamma\left(1-i \kappa_{i,j}\right)+\arg b_{i,j}\right)} (2 \eta)^{\pm i\kappa_{i,j}}  \beta_{i,j}^{\pm 1},\\
    \beta_{i,j} =&  (4v)^{-i\kappa_{i,j}}e^{i\frac{\eta}{2}\frac{(a_j-a_i)^2}{2v}}  \prod_{k=1}^{i-1}p^{1/2}_{k,j}\prod_{l=N+1}^{j-1}p_{i,l}^{-1/2}\nonumber\\
    &\times e^{-i\left(\sum\limits_{\substack{k=1\\k\neq i}}^{N}\kappa_{k,j}\log|a_k-a_i|+\sum\limits_{\substack{l=N+1\\l\neq j}}^{2N}\kappa_{i,l}\log |a_l-a_j|\right)},
\end{align}
where $I_n$ is an $n\times n$ identity matrix and $p_{i,j}$ is called the LZ probability. We note that $M_{i,j}$ is not unitary, and the unitarity of the S-matrix is not manifest. Furthermore, when the off-diagonal elements are large, and the LZ probability is small, the term $p^{-1/2}_{i,j}$ included in $\tilde N^{(-)}$ becomes large. Such a term is likely to cause errors in numerical calculations, so this formula would not be suited for numerical calculations for such cases. We also note that the reality condition is not satisfied in this model. The above derivation of the S-matrix, however, is not problematic. This is because the Hamiltonian with $a_i$ shifted infinitesimally as $a_i+\epsilon_i$ for $i<N$ and $a_{i-N}$ shifted infinitesimally as $a_{i-N}+\epsilon_i$ for $i>N$, where $\epsilon_i\neq \epsilon_j$ for all $i,j$, satisfies the reality condition, while it is obvious that the S-matrix varies continuously with respect to $\epsilon_i$.

Actually, however, we observe that the above matrix \eqref{eq:prev_S} can be transformed into a product of unitary matrices $U_{i,j}$ as follows (GAIA, see Appendix \ref{sec:app1}  for its derivation):
\begin{align}
    S=& \tilde U_{1}\cdots\tilde U_{2N-2}\tilde U_{2N-1}, \\
    \tilde U_{k}=&\prod_{\substack{j-i=k \\ 1\leq i\leq N<N+1\leq j\leq 2N}}U_{i,j},\label{eq:S-matrix}\\
    U_{i, j}=&\left(\begin{array}{ccccc}
    I_{i-1} & & & & \\
     & p^{1/2}_{i ,j} & \cdots & -\tilde\alpha_{i ,j}^{+} & \\
     & \vdots & I_{j-i-1} & \vdots & \\
     & -\tilde \alpha_{i ,j}^{-} & \cdots & p^{1/2}_{i ,j} & \\
     & & & & I_{2 N-j}
    \end{array}\right),\\
    \tilde \alpha_{i ,j}^{\pm}=&\pm \sqrt{1-p_{i,j}}\\
    &\times e^{\pm i\left( \pi / 4+ \arg \Gamma\left(1-i \kappa_{i,j}\right)+\arg b_{i,j}\right)} (2 \eta)^{\pm i\kappa_{i,j}} \tilde  \beta_{i,j}^{\pm 1},\\
    \tilde \beta_{i,j} =&  (4v)^{-i\kappa_{i,j}}e^{i\frac{\eta}{2}\frac{(a_j-a_i)^2}{2v}}\nonumber \\
    &\times e^{-i\left(\sum\limits_{\substack{k=1\\k\neq i}}^{N}\log|a_k-a_i|^{\kappa_{k,j}}+\sum\limits_{\substack{l=N+1\\l\neq j}}^{2N}\log |a_l-a_j|^{\kappa_{i,l}}\right)}.
\end{align}
Hereafter, we express $\tilde \alpha_{i ,j}^{\pm}$ as
\begin{align*}
    \tilde \alpha_{i ,j}^{\pm}=&\pm \sqrt{1-p_{i,j}}e^{\pm i\theta_{i,j}},\\
    \theta_{i,j}=&\frac{ \pi}{4}+ \arg \Gamma\left(1-i \kappa_{i,j}\right)+\arg b_{i,j}\\
    &+\frac{\eta}{2}\frac{(a_j-a_i)^2}{2v}+\log\left(\frac{2 \eta}{4v}\right)^{\kappa_{i,j}} -\Theta_{i,j},\\
    \Theta_{i,j}=&\sum_{\substack{k=1\\k\neq i}}^{N}\log|a_k-a_i|^{\kappa_{k,j}}+\sum_{\substack{l=N+1\\l\neq j}}^{2N}\log |a_l-a_j|^{\kappa_{i,l}}.
\end{align*}
We note that the above derivation is not a direct generalization of AIA to multilevel systems. However, this formula avoids the difficulty of generalizing the AIA directly to multilevel systems. We also note that the S-matrix described above is sufficient to determine the transition probabilities between the basis states of the matrix. To consider a superposition of these states as an initial state, however, it is necessary to take into account the adiabatic evolutions from the initial time $t=-\infty$ to the first anticrossing and from the last anticrossing to the final time $t=\infty$. We will discuss these points in Sec. \ref{subsec:comparison_AIA}.

In addition, this transformation not only makes unitarity clear and calculation simple, but also eliminates the need for a normalization factor in $t\to\pm\infty$. This implies that the formula can be applied to time-periodic systems, for example, which do not have a limit at $t\to\pm\infty$. This will be discussed in Sec. \ref{sec:LZSM}.

Now, we consider the physical meaning of $\eta$. Since the diagonal elements of the Hamiltonian contain terms $\pm\eta vt$, we define a dimensionless time parameter $\tau=\sqrt{\eta v} t$ and a dimensionless Hamiltonian $\tilde H(\tau)=H/\sqrt{\eta v}$. The diagonal elements of $\tilde H$ are of order $O(\eta^0)$, while the off-diagonal elements become $b_{i,j}/\sqrt{v}$. Therefore, the time interval of anticrossing is $\Delta \tau= O(\sqrt{\eta/v})$, while the time interval in which the LZ transition occurs is $\Delta\tau_{LZ}\sim\max\{1,\sqrt{\kappa_{i,j}}\}/\sqrt{2}$~\cite{Shevchenko2010-ic}. If $\eta$ is large enough, $\Delta\tau\gg\Delta\tau_{LZ}$ holds and the LZ transition at each anticrossing can be regarded as independent. This does not mean, however, that the local time evolution $\tilde U_{i,j}$ can be written with only local parameters such as $\kappa_{i,j}$ when $\eta$ is sufficiently large. In fact, $\Theta_{i,j}$ contains a non-local contribution, and we will see in the next section that this term makes an essential contribution to the transition probabilities.

\subsection{Example\label{subsec:ex1}}
\begin{figure}[H]
    \centering
    \includegraphics[width=60mm]{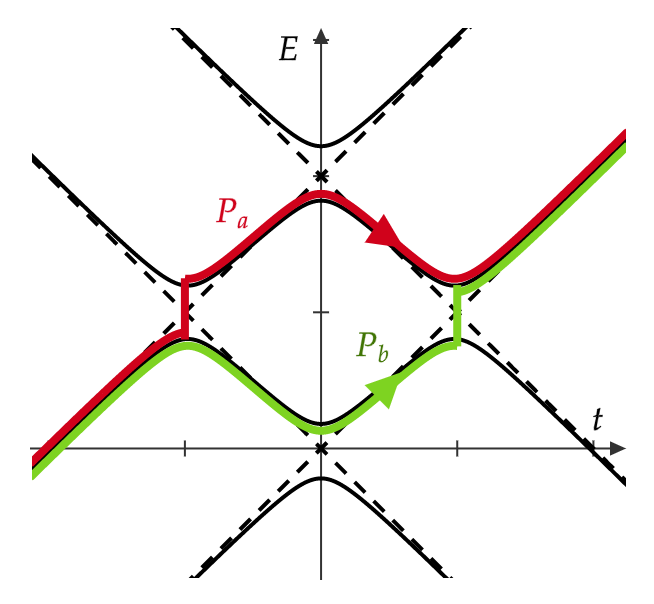}
    \caption{Paths corresponding to $P_a$ and $P_b$. The third term of \eqref{eq:p34} results from interference of these paths.}
    \label{fig:energypath}
\end{figure}

\begin{widetext}
We consider a double quantum dot system to confirm the validity of the S-matrix \eqref{eq:S-matrix}. This system has been experimentally realized and its dynamics has drawn much attention~\cite{Mi2018-rf,Ginzel2020-sl}. Various methods have been proposed to derive an approximate solution for the dynamics~\cite{yurovsky2001curve,malla2017loss,malla2021nonadiabatic}.

Here, we consider the Hamiltonian with $N=2$, that is, a four-level system. In the following, we set $b_{13}=b_{24}=\Delta$ and $b_{14}=b_{23}=\gamma$. In this case, the S-matrix reads as follows:
\begin{align}
    S=\left(\begin{array}{cccc}
    \sqrt{p_{14} p_{13}} & 0 &- \tilde\alpha_{13}^{+}  & -\sqrt{p_{13}}\tilde\alpha_{14}^{+}  \\
    \tilde{\alpha}_{24}^{+} \tilde{\alpha}_{14}^{-} \sqrt{p_{23}}+\tilde{\alpha}_{23}^{+} \tilde{\alpha}_{13}^{-} \sqrt{p_{14}} & \sqrt{p_{23} p_{24}} & -\tilde\alpha_{23}^{+}  \sqrt{p_{13}} & -\tilde{\alpha}_{24}^{+} \sqrt{p_{23} p_{14}}-\tilde{\alpha}_{23}^{+} \tilde{\alpha}_{13}^{-} \tilde{\alpha}_{14}^{+} \\
    -\tilde{\alpha}_{23}^{-} \tilde{\alpha}_{24}^{+} \tilde{\alpha}_{14}^{-}-\tilde{\alpha}_{13}^{-} \sqrt{p_{23} p_{14}} & -\tilde\alpha_{23}^{-}  \sqrt{p_{24}} & \sqrt{p_{23} p_{13}} & \tilde{\alpha}_{24}^{+} \tilde{\alpha}_{23}^{-} \sqrt{p_{14}}+\tilde{\alpha}_{13}^{-} \tilde{\alpha}_{14}^{+} \sqrt{p_{23}} \\
    -\sqrt{p_{24}}\tilde\alpha_{14}^{-} & -\tilde\alpha^{-} _{24} & 0 & \sqrt{p_{24} p_{14}}
    \end{array}\right).\label{eq:S_4}
\end{align}

We will compare the approximate solution \eqref{eq:S_4} obtained by GAIA with the solution obtained by numerical calculation with the Python Library QuTiP~\cite{Johansson2012-du,Johansson2013-gw}. First, we confirm that GAIA is a good approximation in the region where the approximation that the LZ transitions occur independently is appropriate (Figure~\ref{fig:GAIA4_eta}(a)). In Figure~\ref{fig:GAIA4_eta}(a), the red background region corresponds to the parameter region where the LZ transitions cannot be considered independent $\sqrt{\eta/v}a\not\gg\sqrt{2}\max_{i,j}\{1,\sqrt{\kappa_{i,j}}\}$. Outside of this region, we can see that the numerical solution is in good agreement with \eqref{eq:S_4}.

\begin{figure}[b]
    \centering
    \begin{overpic}[width=60mm]{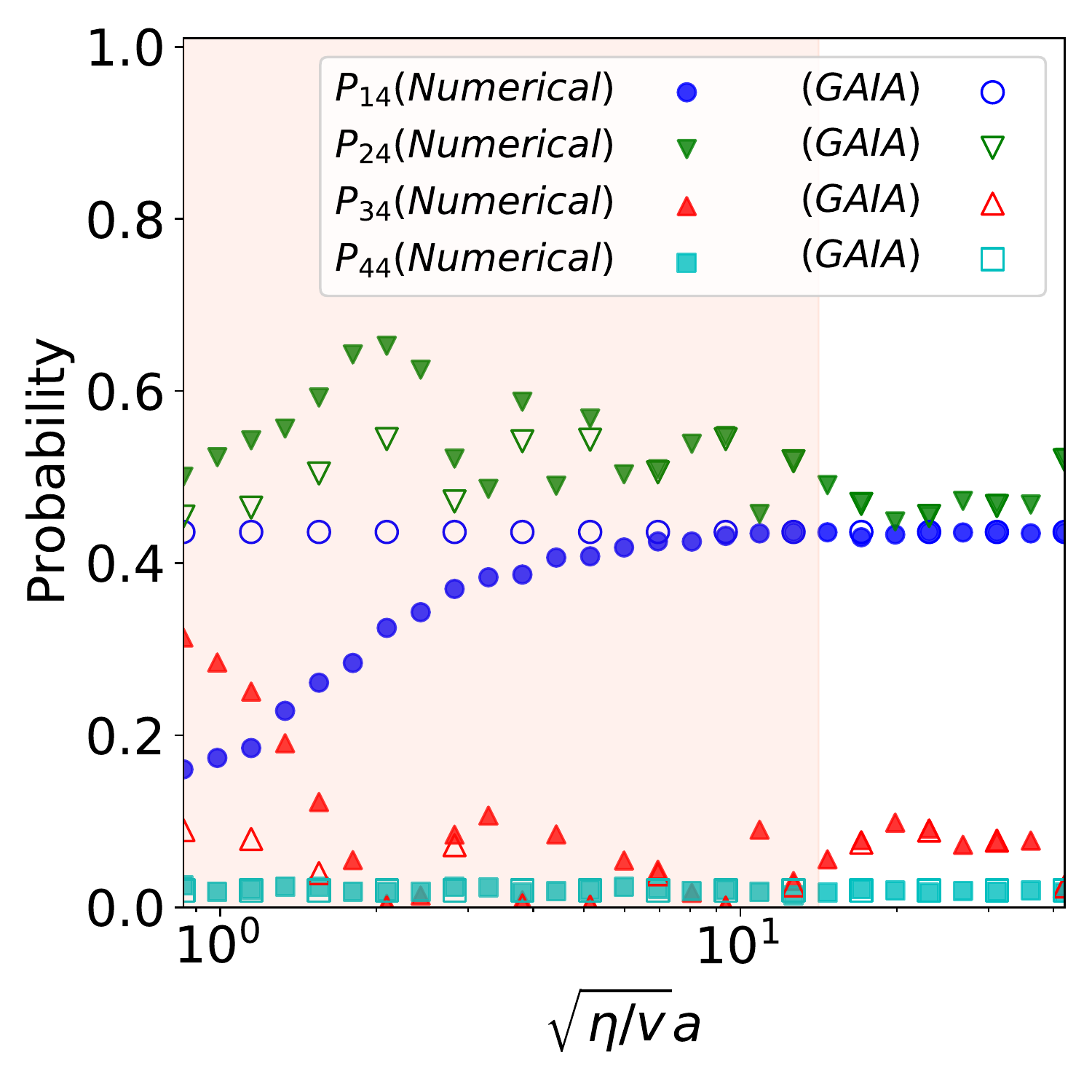}
        \put(19,90){\Large{(a)}}
    \end{overpic}
    \begin{overpic}[width=60mm]{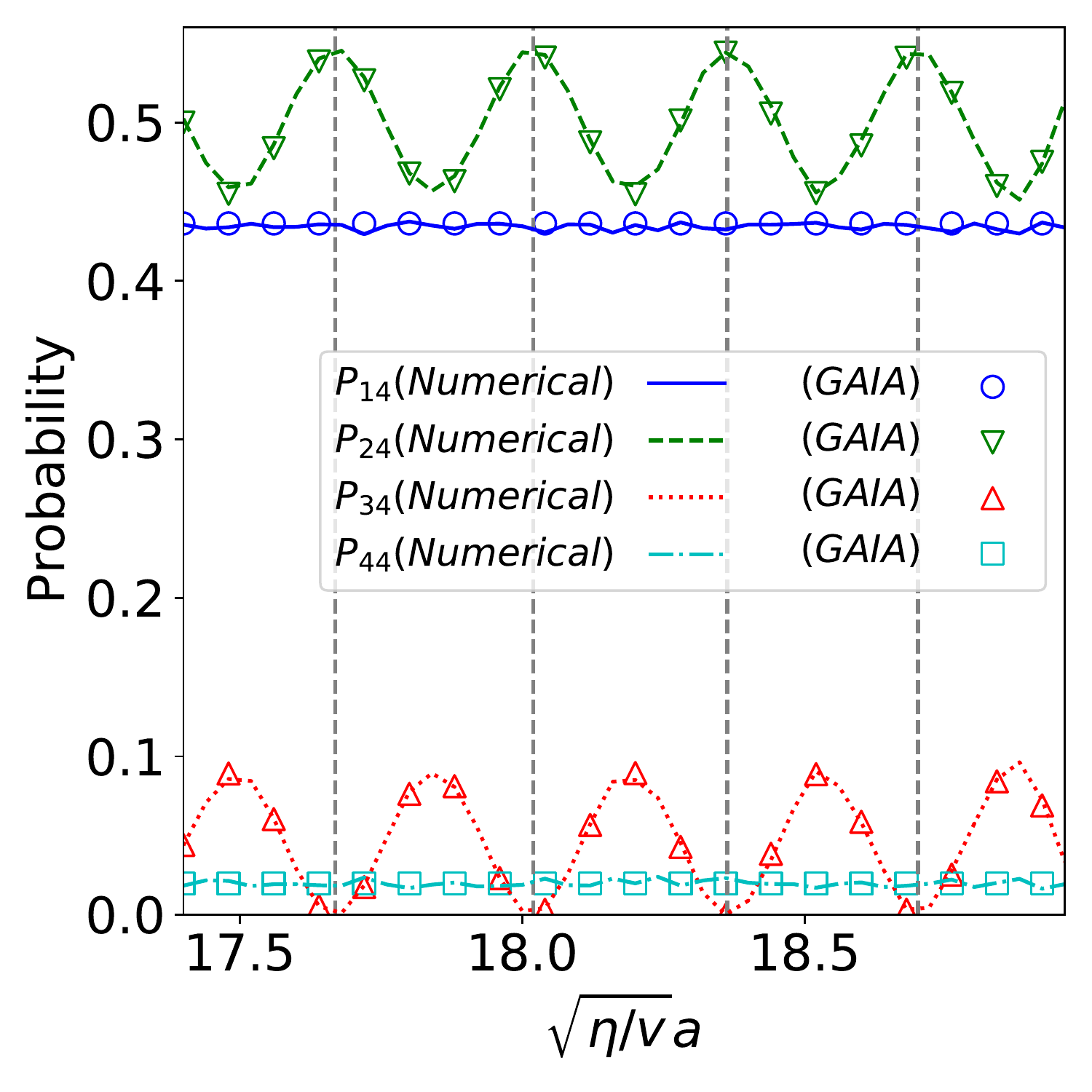}
        \put(19,90){\Large{(b)}}
    \end{overpic}
    \caption{(a) Numerical results of the transition probability from the initial state $\ket{4}$ (closed symbol) and plot of $P_{i4}=|S_{i4}|^2$ in~\eqref{eq:S_4} obtained by GAIA (open symbol). The parameters are set to $\Delta/\sqrt{v}=0.5$ and $\gamma/\sqrt{v}=1.0$. The parameter region with red background is where the LZ transition cannot be considered independent : $\sqrt{\eta/v}a<10\sqrt{2}\gamma/\sqrt{v}$. As the parameter region is reached where the LZ transitions can be regarded as independent, the numerical and GAIA results coincide. (b) Enlarged view of (a). The gray dashed vertical lines indicate the points where $P_{34}$ \eqref{eq:p34} vanishes. We used the solver for numerical calculation implemented in the Python Library QuTiP~\cite{Johansson2012-du,Johansson2013-gw}.}
    \label{fig:GAIA4_eta}
\end{figure}

Next, we look at the contribution from the non-local term $\Theta_{i,j}$. To do so, consider $P_{34}=|S_{34}|^2$, which is the probability of measuring $\ket{3}$ when the initial state is $\ket{4}$ and is given by

\begin{equation}
    P_{34}=P_a+P_b+2\sqrt{ P_a P_b}\cos \biggl(2\arg \left(\frac{\Gamma\left(1-i \kappa_{\gamma}\right) }{\Gamma\left(1-i \kappa_{\Delta}\right)}\right)+\frac{\eta a^{2}}{2 v}+2\log \left(\frac{\eta a^{2}}{2 v}\right)^{-\kappa_{\Delta}+\kappa_{\gamma}}\biggr),\label{eq:p34}
\end{equation}
\end{widetext}
where we define $ P_a:=p_{14}\left(1-p_{24}\right)\left(1-p_{23}\right)$, $ P_b:=\left(1-p_{14}\right)\left(1-p_{13}\right) p_{23}$, $\kappa_\Delta:=\kappa_{13}=\kappa_{24}$, and $\kappa_\gamma:=\kappa_{14}=\kappa_{23}$. The $P_a$ and $P_b$ correspond to the transition probabilities on the paths in Figure~\ref{fig:energypath}, and the third term is their interference term. The non-local term $\Theta_{i,j}$ contributes to the last term of the phase. The zeros of \eqref{eq:p34} are shown in Figure~\ref{fig:GAIA4_eta}(b) as gray dashed lines. We can see that the zeros of \eqref{eq:p34} and the zeros of the transition probabilities calculated numerically are in good agreement. This result shows that even in the parameter region where the LZ transitions can be regarded as independent, the non-local terms make an essential contribution.

\subsection{Comparison with AIA\label{subsec:comparison_AIA}}

\begin{figure}[t]
    \centering
    \includegraphics[width=90mm]{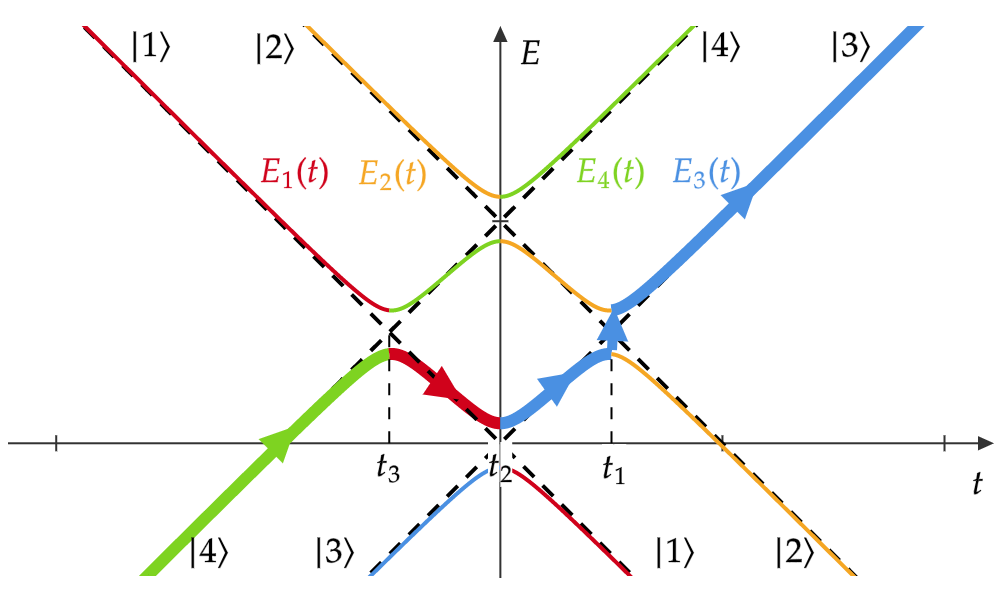}
    \caption{The correspondence between a path and a transition amplitude. The figure shows a path $(t_I,t_{3};E_{4})\to(t_{3},t_{2};E_{1})\to(t_{2},t_{1};E_{3})\to (t_1,t_F;E_{3})$. The time evolution is considered adiabatic at $t\neq t_i$ in AIA. If the LZ transition happens at $t=t_i$, we multiply the transition amplitude by LZ transition probability. On the other hand, if the LZ transition does not happen at $t=t_i$, we multiply the transition amplitude by $\hat\alpha^\pm$.}
    \label{fig:energypath_corresponding}
\end{figure}

In this subsection, we investigate the relationship between AIA, an approximation method that has been used so far, and GAIA proposed here. There are two types of AIAs: those using energy basis $\ket{\epsilon_i(t)}$~\cite{Shevchenko2010-ic,Niranjan2020-ta} and those using diabatic basis $\ket{i}$~\cite{Kayanuma1997-ec,Ostrovsky2007-dg}. Both approximation methods use instantaneous eigenvalues to describe the adiabatic time evolution, which has a disadvantage of being difficult to compute analytically for multilevel systems. Here, we consider a formulation using diabatic basis. In the following, for simplicity, let $a_n=(n-1)a\ (1\leq n\leq N)$. In this case, AIA yields the following S-matrix~\cite{Kayanuma1997-ec}.

\begin{align}
    S=&  K_{1}\tilde G_{1}\cdots \tilde G_{2N-2} K_{2N-1}\tilde G_{2N-1} K_{2N}, \label{eq:S_AIA}\\
    \tilde G_k=& \prod_{\substack{j-i=k \\ 1\leq i\leq N, N+1\leq j\leq 2N}}G_{i,j}, \\
    G_{i,j}=&\left(\begin{array}{ccccc}
    I_{i-1} & & & & \\
     & p_{i ,j}^{1/2} & \cdots & -\hat \alpha^{+}_{i,j} & \\
     & \vdots & I_{j-i-1} & \vdots & \\
     & -\hat \alpha^{-}_{i,j} & \cdots & p_{i,j}^{1/2} & \\
     & & & & I_{2 N-j}
    \end{array}\right),\\
    K_{k}=&e^{-i\int_{t_{k-1}}^{t_k}dt\operatorname{diag}\left(
    E_1(t),E_2(t),\cdots,E_{2N}(t)\right)},\\
    \hat \alpha^{\pm}_{i,j}=&\pm(1-p_{i,j})^{1/2}e^{\pm i\varphi_{i,j}},\\
    \varphi_{i ,j}=&\frac{\pi}{4}+\kappa_{i ,j}(\ln \kappa_{i ,j}-1)+\arg \Gamma(1-i \kappa_{i ,j}),
\end{align}
where $t_k$ stands for the $(2N-k)$th anticrossing time and $E_i(t)$, which is related to $i$th diagonal element $H_{ii}(t)$ of Hamiltonian, is the instantaneous eigenvalue $\epsilon_n(t)$ in the time interval $[t_k,t_{k-1}]$ (Figure~\ref{fig:energypath_corresponding}): for example, for $1\leq i\leq N$
\begin{align}
    E_i(t)=
    \begin{dcases}
    \epsilon_{N+i}(t)&t\leq t_{2N-i}\\
    \epsilon_{N-k+2i-1}(t)&t_{2N-k}\leq t\leq t_{2N-k-1}\\
    &(k=i,\cdots, i+N-2)\\
    \epsilon_{i}(t)& t_{N+i-1}<t
    \end{dcases},
\end{align}
and for $N+1\leq i\leq 2N$
\begin{align}
    E_i(t)=
    &\begin{dcases}
    \epsilon_{i-N}(t)&t\leq t_{i-1}\\
    \epsilon_{2i-l-N}(t)&t_{l}\leq t\leq t_{l-1}\\
    &(l=i+1-N,\cdots,i-1)\\
    \epsilon_{i}(t)& t_{l-N}<t
    \end{dcases}.
\end{align}

\begin{widetext}
The S-matrix (\ref{eq:S_AIA}) maps one transition amplitude to one path in the energy diagram (Figure~\ref{fig:energypath_corresponding}). For example, if we specify a path as $(t_I,t_F;E_i)$, then for a path
\begin{align*}
    (t_I,t_{2N-1};E_{i_1})&\to(t_{2N-1},t_{2N-2};E_{i_2})\to\cdots\to (t_1,t_F;E_{i_{2N}}),
\end{align*} 
a transition amplitude
\begin{align}
    e^{-i\int^{t_F}_{t_1}dt E_{i_{2N}}(t)} \cdots (G_{i_2,i_3})_{i_3,i_2}e^{-i\int^{t_{2N-2}}_{t_{2N-1}}dt E_{i_2}(t)} (G_{i_1,i_2})_{i_2,i_1} e^{-i\int^{t_{2N-1}}_{t_{I}}dt E_{i_1}(t)}
\end{align}
is selected, where $t_I$ and $t_F$ are initial and final times. Then, before and after an anticrossing, we can rewrite the amplitude as
\begin{align}
    &e^{-i\int^{t_{2N-k-2}}_{t_{2N-k-1}}dt E_{i_{k+2}}(t)} (G_{i_{k+1},i_{k+2}})_{i_{k+2},i_{k+1}}e^{-i\int^{t_{2N-k-1}}_{t_{2N-k}}dt E_{i_{k+1}}(t)}\nonumber\\
    =&e^{-i\int^{t_{2N-k-2}}_{t_{R}}dt E_{i_{k+2}}(t)}\bigl((G_{i_{k+1},i_{k+2}})_{i_{k+2},i_{k+1}}e^{-i\int^{t_{2N-k-1}}_{t_{R}}dt(E_{i_{k+1}}(t)-E_{i_{k+2}}(t))}\bigr)e^{-i\int^{t_{R}}_{t_{2N-k}}dt E_{i_{k+1}}(t)},
\end{align}
\end{widetext}
where $t_R$ is an arbitrary time. Using this transformation, the S-matrix \eqref{eq:S_AIA} can be transformed as follows:
\begin{align}
    S=&K'_{1}\tilde G'_{1}\cdots \tilde G'_{2N-2} \tilde G'_{2N-1}K'_{2N}, \label{eq:S_AIA_revise}\\
    \tilde G'_k=& \prod_{\substack{j-i=k \\ 1\leq i\leq N, N+1\leq j\leq 2N}}G_{i,j}^{(k)}, \\
    G_{i,j}^{(k)}=&\left(\begin{array}{ccccc}
    I_{i-1} & & & & \\
     & p_{i ,j}^{1/2} & \cdots & -\hat{\tilde \alpha}^{(k)+}_{i,j}  & \\
     & \vdots & I_{j-i-1} & \vdots & \\
     &-\hat{\tilde \alpha}^{(k)-}_{i,j}  & \cdots & p_{i,j}^{1/2} & \\
     & & & & I_{2 N-j}
    \end{array}\right),\\
    \hat{\tilde \alpha}^{(k)\pm}_{i,j}=&\pm(1-p_{i,j})^{1/2}e^{\pm i\tilde \varphi^{(k)}_{i,j}},\\
    \tilde \varphi^{(k)}_{i ,j}=&\frac{\pi}{4}+\kappa_{i ,j}(\ln \kappa_{i ,j}-1)+\arg \Gamma(1-i \kappa_{i ,j})\\
    &+\int_{t_{R}}^{t_{k}}dt( E_{i}(t)-E_j(t)), \label{eq:int_energy}\\
    K_{1}'=&e^{-i\int^{t_{F}}_{t_{R}}dt\operatorname{diag}(E_1(t),E_2(t),\cdots,E_{2N}(t))},\\
    K_{2N}'=&e^{-i\int^{t_{R}}_{t_{I}}dt\operatorname{diag}(E_1(t),E_2(t),\cdots,E_{2N}(t))}.
\end{align}
We note here that the contribution from $t_R$ can be neglected since it offsets the other terms. If we ignore $t_R$, we can see that the S-matrix \eqref{eq:S_AIA_revise} is similar to the S-matrix of GAIA \eqref{eq:S-matrix}. As we mentioned before, when computing the transition amplitude for the case where the initial state is a superposition of eigenstates, the S-matrix \eqref{eq:S-matrix} is not correct, and the contribution of $K'_{2N}$ and $K'_{1}$ in \eqref{eq:S_AIA_revise} must be taken into account. 

It can be seen that the S-matrix~\eqref{eq:S_AIA_revise} obtained in AIA is similar to the previously obtained S-matrix~\eqref{eq:S-matrix}. If we perturbatively expand the integral of $E_i(t)-E_j(t)$ formally, we get
\begin{align*}
    \int_{t_R}^{t_{k}}dt( E_{i}(t)-E_j(t))=&\int_{t_R}^{t_{k}}dt (H_{ii}(t)-H_{jj}(t))\\
    &+\int_{t_R}^{t_{k}}dt \left(\sum_{n=1}^{N}\frac{-\eta|b_{n,j}|^2}{ H_{jj}(t)-H_{nn}(t)}\right.\\
    &\quad \left.+\sum_{m=N+1}^{2N}\frac{\eta|b_{i,m}|^2}{ H_{ii}(t)-H_{mm}(t)}\right)\\
    &+O(\eta^{-1})
\end{align*}
because $H_{ii}(t)\propto \eta$. This perturbation, however, breaks down at the anti-crossing point since we have $H_{nn}(t)-H_{mm}(t)=-2\eta v (t-t_{m-n})\ (1\leq n\leq N,N+1\leq m\leq2N)$. It can be seen that GAIA successfully avoids the breakdown of the perturbation of AIA. Therefore, the way of obtaining the S-matrix~\eqref{eq:S-matrix} previously is a more general one that extends the AIA to multilevel systems.

\section{Multilevel Landau--Zener--St\"uckelbelg--Majorana Interference Model\label{sec:LZSM}}
\subsection{GAIA\label{subsec:gaia_for_LZSM}}

In this section, we consider the multilevel Landau--Zener--St\"uckelberg--Majorana (LZSM) interference model, which is an extension of the LZSM interference model of two-level systems~\cite{Shevchenko2010-ic} and includes, for example, the photon-assisted Landau-Zener model, which is a system of a driven spin and single-mode boson~\cite{werther2019davydov,sun2012photon,Lidal2020-go,Wang2021-fy,Zheng2021-qb,Neilinger2016-jw,Bonifacio2020-lr}.
The model is described by the Hamiltonian
\begin{align*}
    H(t)=&\eta A(t)+\sqrt{\eta}B,\\
    A(t)=&\operatorname{diag}(-\sin(vt)+a_1,\ldots,-\sin(vt)+a_N,\\
    &\quad\quad\quad\sin(vt)+a_1,\ldots,\sin(vt)+a_N),\\
    B=&\begin{pmatrix}
    0&\cdots&0&b_{1,N+1}&\cdots&b_{1,2N}\\
    \vdots&\ddots&\vdots&\vdots&\ddots&\vdots\\
    0&\cdots&0&b_{N,N+1}&\cdots&b_{N,2N}\\
    b_{1,N+1}^\ast&\cdots&b_{N,N+1}^\ast&0&\cdots&0\\
    \vdots&\ddots&\vdots&\vdots&\ddots&\vdots\\
    b_{1,2N}^\ast&\cdots&b_{N,2N}^\ast&0&\cdots&0
    \end{pmatrix}.
\end{align*}

We consider the reality condition, i.e., a condition for the existence of energy gaps without anticrossing~\cite{Aoki2002-zw,shimada2020numerical}. The existence of such gaps requires the consideration of new Stokes curves and the discussion in the previous section cannot be used. The reality condition is satisfied when the relation
\begin{align}
    \frac{\eta}{H_{ii}(t)-H_{jj}(t)} &=\sum_{m} \frac{1}{\lambda_{i,j, m}\left(t-t_{i,j, m}\right)} \label{eq:pole_condition}\\
    \lambda_{i,j, m} &=\frac{1}{\eta}\left.\frac{d}{d t}\left(H_{jj}(t)-H_{ii}(t)\right)\right|_{t=t_{i,j, m}}
\end{align}
holds, where $t_{i,j,m}$ stands for the $m$th zero of the denominator (i.e, the time when the $i$th state and the $j$th state cross in the energy diagram for the $m$th time). We note that we do not impose the positivity of $\lambda_{i,j,m}$ here, but we fix $1\leq i\leq N$ and $N+1\leq j\leq2N$. Hereafter, we impose $\lambda_{i,j,m}\neq 0$ if $b_{i,j}\neq0$. The condition~\eqref{eq:pole_condition} is always satisfied whenever the left-hand side is analytic (including infinity) except for simple poles on the real $t$. In the multilevel LZSM interference model, indeed, if $|a_i-a_j|<2 \ (1\leq i\leq N,N+1\leq j\leq2N)$ is satisfied,
\begin{align}
    &\frac{\eta}{H_{ii}(t)-H_{jj}(t)}\\
    =&\sum_{n}\left((-1)^n2v\sqrt{1-\frac{(a_i-a_j)^2}{4}}\right.\\
    &\times\left.\left(t+\frac{(-1)^{n}\arcsin\left(\frac{a_i-a_j}{2}\right)+(n-1)\pi}{v}\right)\right)^{-1}
\end{align}
holds, so the reality condition is satisfied.

Since the instantaneous eigenstates of the Hamiltonian do not coincide with the computational basis at $t=\pm\infty$, the method of calculating the S-matrix in the previous study~\cite{Aoki2002-zw,shimada2020numerical} cannot be used for this model. Formally, $M_{i,j}$ can be obtained, but the normalization factor $\tilde{N}^{(\pm)}$ cannot be obtained. On the other hand, in the previous section, we have seen that the  same S-matrix is obtained just by replacing the matrix $M_{i,j}$ with the unitary matrix $U_{i,j}$ and discarding the normalization factors  $\tilde{N}^{(\pm)}$. We therefore consider, also  in this model, a product of the unitary matrices $U_{i,j,n}$ describing the unitary evolution between $|i\rangle$ and $|j\rangle$ across the $n$th anticrossing, in place of  $M_{i,j,n}$ and assume that it represents the S-matrix. Here, the unitary matrix $U_{i,j,n}$ can be written as follows:
\begin{widetext}
\begin{align*}
    U_{i, j,n}=&
    \left(\begin{array}{ccccc}
    I_{i-1} & & & & \\
     & p^{1/2}_{i ,j,n} & \cdots & -\tilde\alpha_{i ,j,n}^{+} & \\
     & \vdots & I_{j-i-1} & \vdots & \\
     & -\tilde \alpha_{i ,j,n}^{-} & \cdots & p^{1/2}_{i ,j,n} & \\
     & & & & I_{2 N-j}
    \end{array}\right),\\
    \tilde \alpha_{i,j,n}^{\pm}=&\pm \operatorname{sgn}(\lambda_{i,j,n})\sqrt{1-p_{i,j,n}}e^{\pm\operatorname{sgn}(\lambda_{i,j,n}) i\theta_{i,j,n}},\quad p_{i,j,n}=e^{-2\pi\kappa_{i,j,n}},\quad \kappa_{i,j,n}=\frac{|b_{ij}|^2}{|\lambda_{i,j,n}|},\\
    \theta_{i,j,n}=&\frac{ \pi}{4}+ \arg \Gamma\left(1-i\kappa_{i,j,n}\right)+\arg b_{i,j}+\zeta_{i,j,n}+\log\left(\frac{ \eta}{|\lambda_{i,j,n}|}\right)^{\kappa_{i,j,n}} -\Theta_{i,j,n},\\
    \Theta_{i,j,n}=&\log\left(\prod_{\substack{m=-\infty\\(m\neq n)}}^{\infty}\left|\lambda_{i,j, m}\left(t_{i,j, n}-t_{i,j, m}\right)\right|^{2\kappa_{i, j, n}}\right.\\
    &\left.\times\prod_{\substack{m=-\infty}}^{\infty}\left(\prod_{\substack{k=1\\k\neq i}}^{N}\left|\lambda_{k, j, m}\left(t_{i,j, n}-t_{k, j, m}\right)\right|^{\kappa_{k, j, m}}\prod_{\substack{l=N+1\\l\neq j}}^{2 N}\left|\lambda_{i, l, m}\left(t_{i,j, n}-t_{i, l, m}\right)\right|^{\kappa_{i, l, m}}\right)\right),\\
    \zeta_{i,j,n}=&\operatorname{sgn}(\lambda_{i,j,n})\int^{t_{(i,j),n}}(H_{ii}(t,\eta)-H_{jj}(t,\eta))dt.
\end{align*}
\end{widetext}
Notice that $\Theta_{i,j,n}$  includes infinite products which, however, can be made simplified:
\begin{align}
    &\prod_{\substack{m=-\infty}}^{\infty}\left|\lambda_{k, j, m}\left(t_{i,j, n}-t_{k, j, m}\right)\right|^{\kappa_{k, j, m}}\\
    &=
    \left|\frac{\sin(\frac{v(t_{i,j, n}-t_{k, j, 0})}{2})}{\sin(\frac{v(t_{i,j, n}-t_{k, j, 1})}{2})}\right|^{\kappa_{k, j, 0}}
\end{align}
and
\begin{align}
    &\prod_{\substack{m=-\infty\\(m\neq n)}}^{\infty}\left|\lambda_{i,j, m}\left(t_{i,j, n}-t_{i,j, m}\right)\right|^{2 \kappa_{i, j, n}}\\
    =&\left|\frac{v}{2\lambda_{i,j, n}\sin(\frac{v(t_{i,j, n}-t_{i, j, n+1})}{2})}\right|^{2 \kappa_{i, j, n}}.
\end{align}
\begin{widetext}
In this way, we need only finite products to calculate $\Theta_{i,j,n}$:
\begin{align}
    \Theta_{i,j,n}
    =&\log\left(\left|\frac{v}{2\lambda_{i,j, n}\sin(\frac{v(t_{i,j, n}-t_{i, j, n+1})}{2})}\right|^{2 \kappa_{i, j, n}}\right.\\
    &\left.\times\prod_{\substack{k=1\\(k\neq i)}}^{N}\left|\tan\left(\frac{v(t_{i,j, n}-t_{k, j, 0})}{2}\right)\right|^{\kappa_{k, j, 0}} \prod_{\substack{l=N+1\\(l\neq j)}}^{2 N}\left|\tan\left(\frac{v(t_{i,j, n}-t_{i, l, 0})}{2}\right)\right|^{\kappa_{i, l, 0}}\right).
\end{align}
\end{widetext}

\subsection{Example}

\begin{figure}[H]
    \centering
    \begin{overpic}[width=57.5mm]{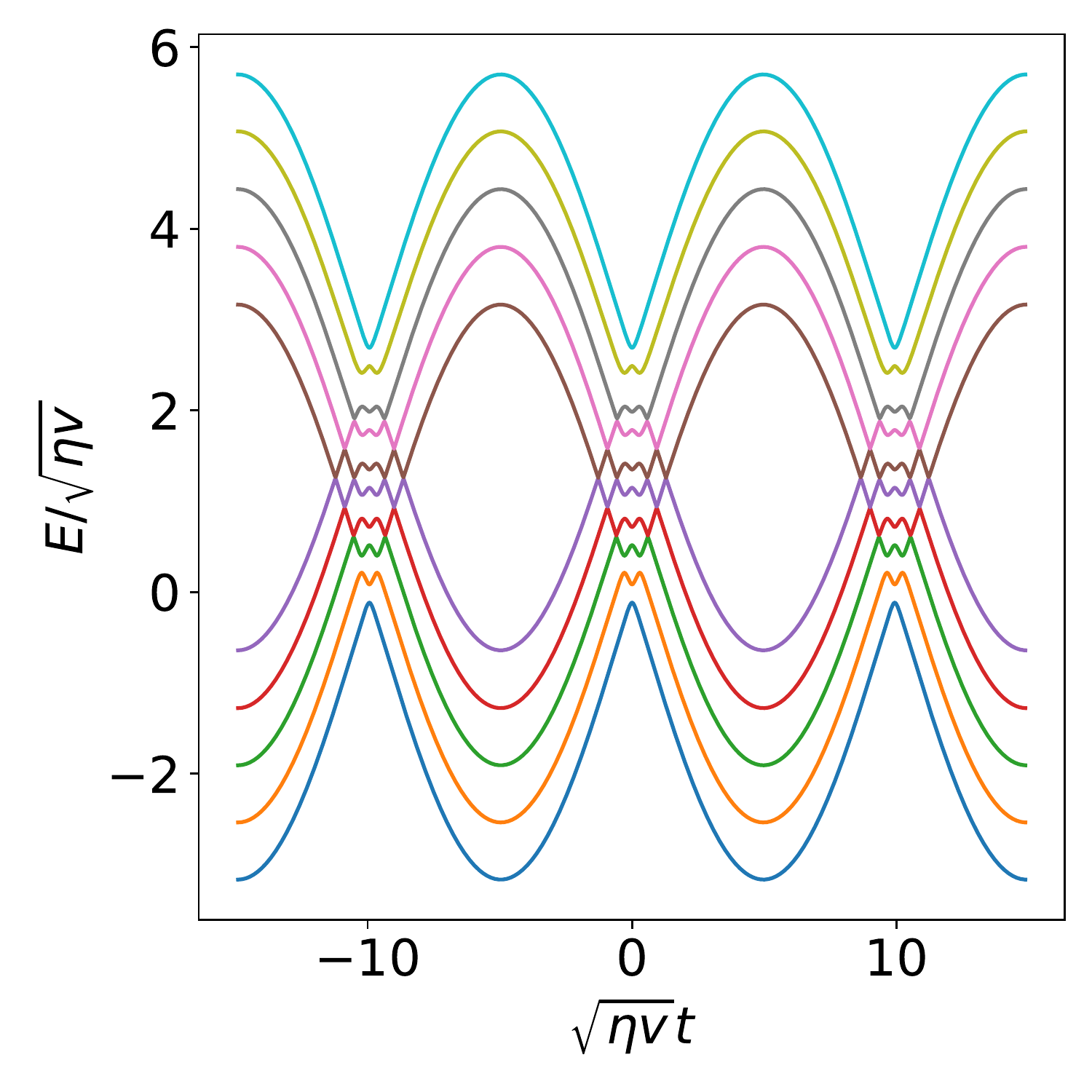}
        \put(19,90){\Large{(a)}}
    \end{overpic}
    \begin{overpic}[width=57.5mm]{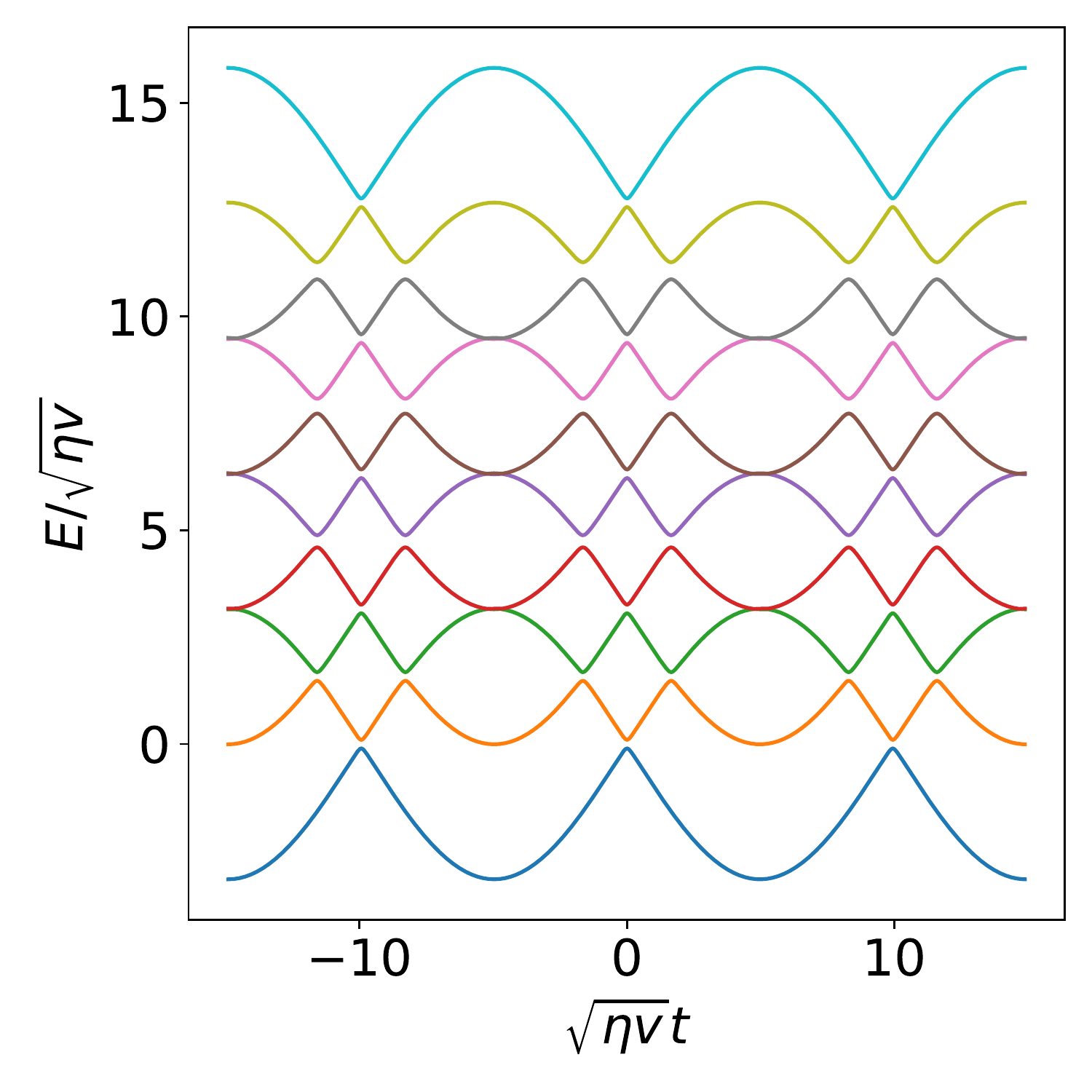}
        \put(19,90){\Large{(b)}}
    \end{overpic}
    \caption{Energy spectra of \eqref{eq:H_spinboson} when $\gamma=0.1\sqrt{v}$, $\Delta=0.1\sqrt{v}$, $\eta=10v$, and (a) $\sqrt{\eta/v}a=0.2$ and (b) $\sqrt{\eta/v}a=1.0$. The dimension of the boson Hilbert space is truncated at $5$. In (b), the transition does not occur at times when the energies are in contact because $b_{i,N+i+3}=0$.}
    \label{fig:energy_LZSM}
\end{figure}

We consider a driven system of a spin and a single-mode boson~\cite{werther2019davydov,sun2012photon,Lidal2020-go,Wang2021-fy,Zheng2021-qb,Neilinger2016-jw,Bonifacio2020-lr}, described by the Hamiltonian 
\begin{align}
    H(t)=&-\eta \sin(v t) \sigma_{z}+\sqrt{\eta}\Delta \sigma_{x}\\
    &+ \eta\Omega b^{\dagger} b+ \sqrt{\eta}\gamma \sigma_{x}\left(b+b^{\dagger}\right)\label{eq:H_spinboson}\\
    =:&\eta A(t)+\sqrt{\eta}B\\
    A(t)=&\operatorname{diag}(-\sin(vt),-\sin(vt)+\Omega,\ldots,\\
    &\quad\quad\quad\sin(vt),\sin(vt)+\Omega,\ldots)\\
    B=&\begin{pmatrix}
    0&0&0&\cdots&\Delta&\gamma&0&\cdots\\
    0&0&0&\cdots&\gamma&\Delta&\sqrt{2}\gamma&\cdots\\
    0&0&0&\cdots&0&\sqrt{2}\gamma&\Delta&\cdots\\
    \vdots&\vdots&\vdots&\ddots&\vdots&\vdots&\vdots&\ddots\\
    \Delta&\gamma&0&\cdots&0&0&0&\cdots\\
    \gamma&\Delta&\sqrt{2}\gamma&\cdots&0&0&0&\cdots\\
    0&\sqrt{2}\gamma&\Delta&\cdots&0&0&0&\cdots\\
    \vdots&\vdots&\vdots&\ddots&\vdots&\vdots&\vdots&\ddots    
    \end{pmatrix}.
\end{align}
Figure~\ref{fig:energy_LZSM} shows the energy spectrum of the Hamiltonian. In the following numerical calculation, the dimension of the boson Hilbert space is truncated at $N=5$. Since the off-diagonal elements are unbounded, they always violate the condition for GAIA $\Delta \tau \gg \Delta\tau_{LZ}$. However, as long as the probability amplitude at the anticrossing that violates the condition for GAIA is $0$, GAIA is considered reasonable.

\begin{figure}[H]
    \centering
    \begin{overpic}[width=65mm]{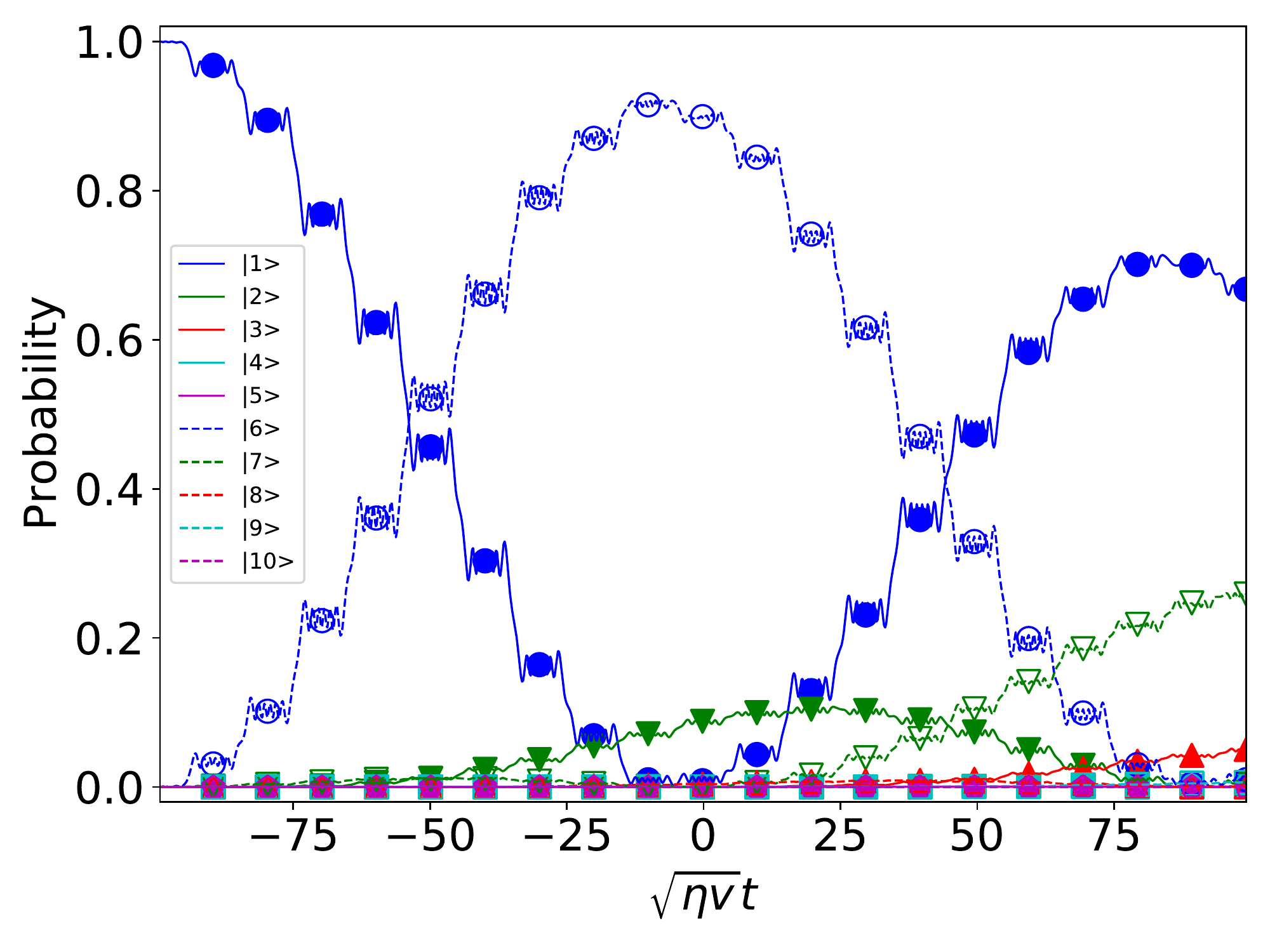}
        \put(19,66.5){\Large{(a)}}
    \end{overpic}
    \begin{overpic}[width=65mm]{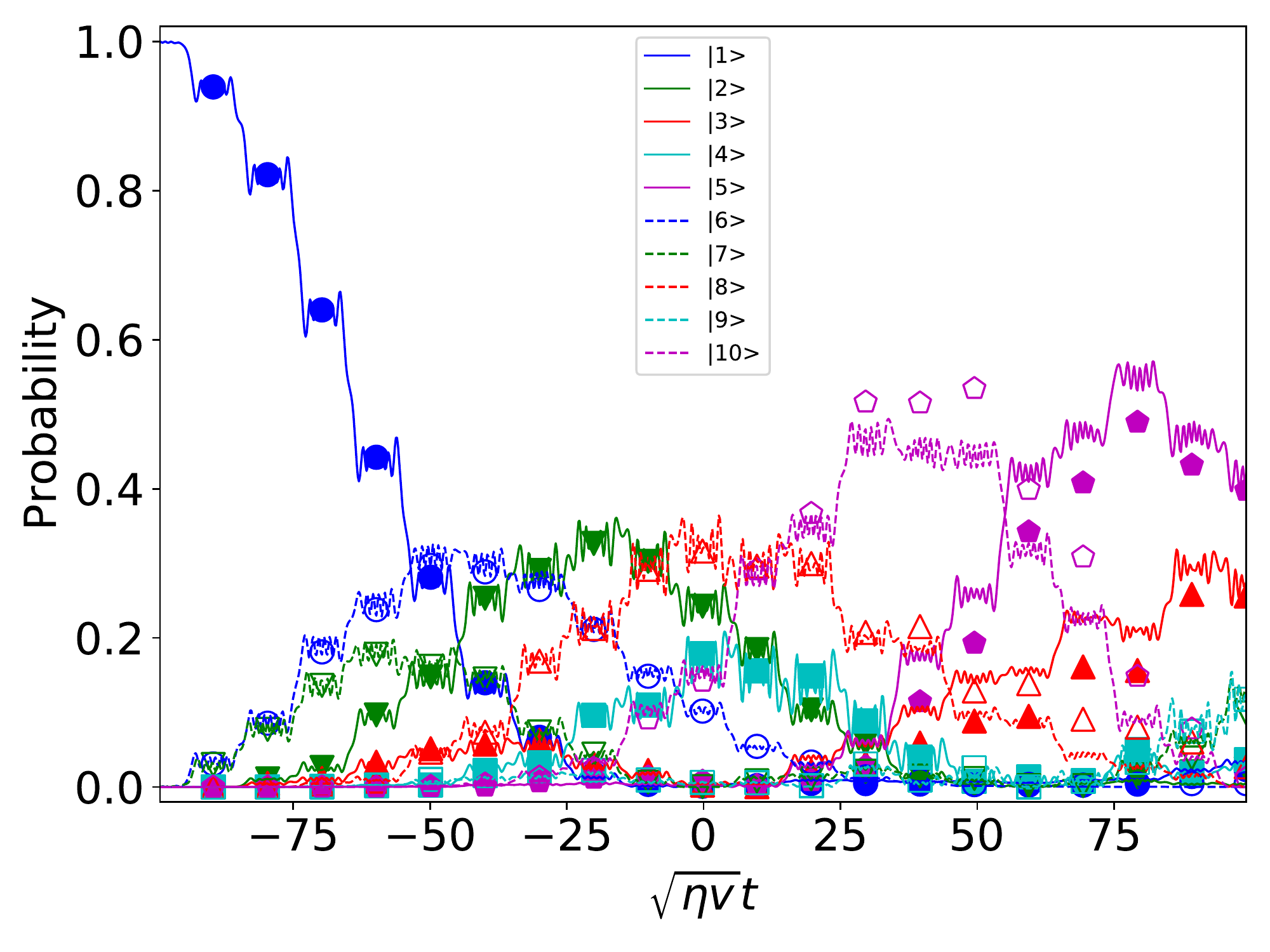}
        \put(19,66.5){\Large{(b)}}
    \end{overpic}
    \begin{overpic}[width=65mm]{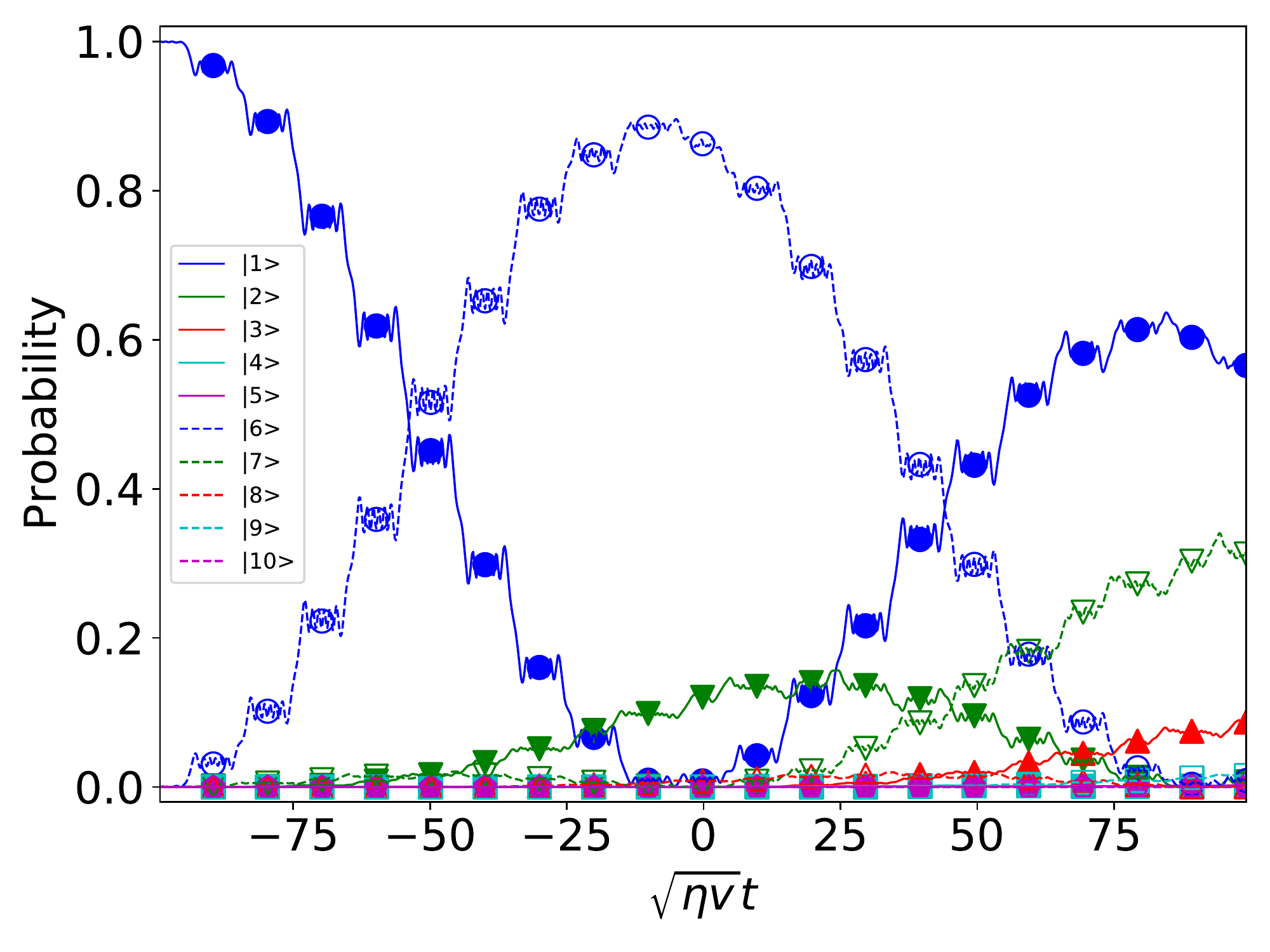}
        \put(19,66.5){\Large{(c)}}
    \end{overpic}
    \begin{overpic}[width=65mm]{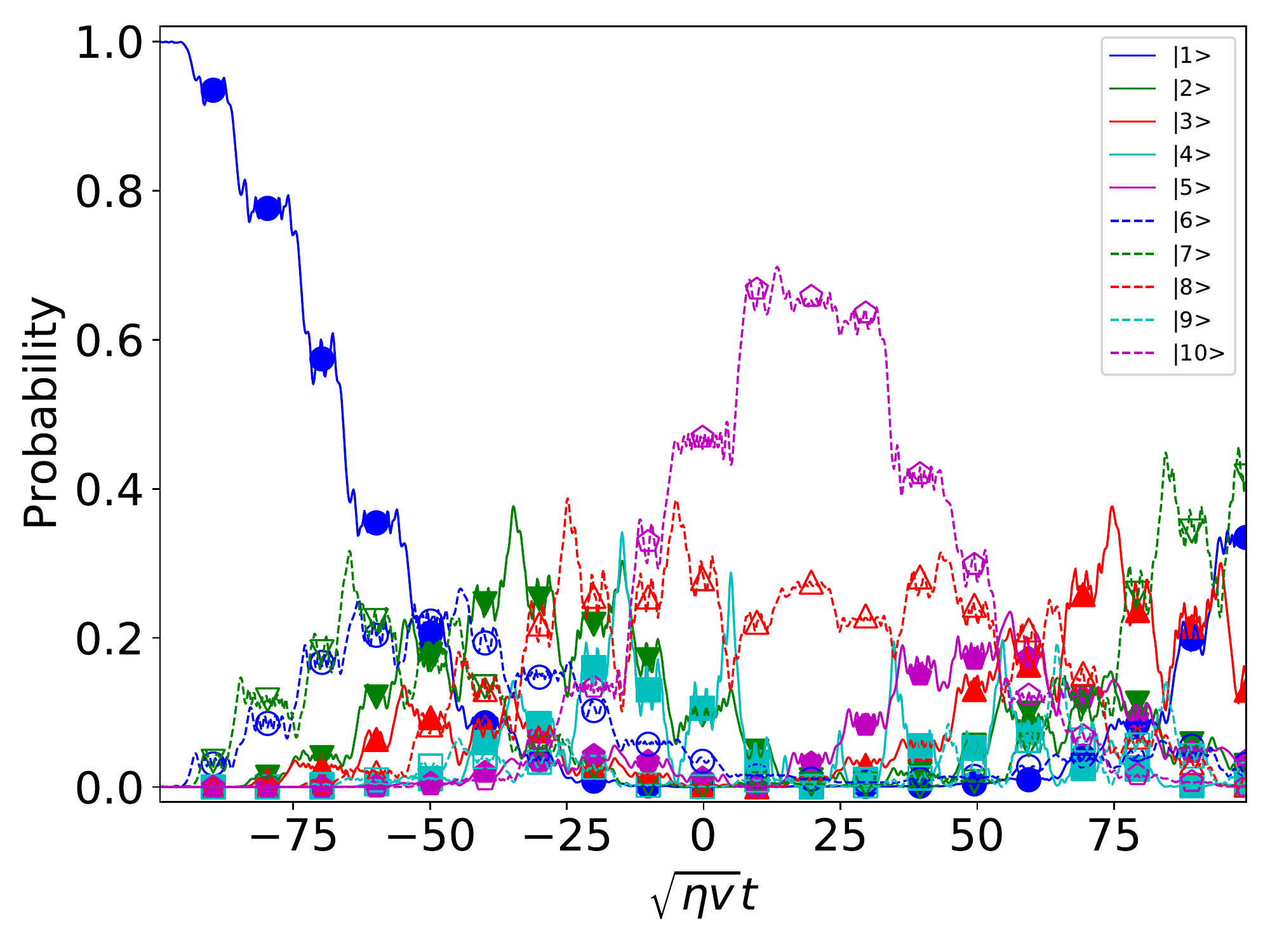}
        \put(19,66.5){\Large{(d)}}
    \end{overpic}
    \caption{Time dependence of the transition probability. The parameters are $\Delta=0.1\sqrt{v}$, $\eta=10v$, and (a) $\sqrt{\eta/v}a=0.2, \gamma=0.02\sqrt{v}$, (b) $\sqrt{\eta/v}a=0.2, \gamma=0.1\sqrt{v}$, (c) $\sqrt{\eta/v}a=1.0, \gamma=0.02\sqrt{v}$, and (d) $\sqrt{\eta/v}a=1.0, \gamma=0.1\sqrt{v}$. Solid lines and dashed lines represent the results of numerical calculations, and the dots represent the results of GAIA.}
    \label{fig:prob_LZSM_02_1}
\end{figure}

The other condition to be satisfied is the reality condition. In the previous discussion, we explained that the reality condition is satisfied in this model by imposing the condition $|a_i-a_j|<2$. Since the reality condition is a condition of the anticrossing, even if the condition $|a_i-a_j|<2$ is not satisfied, GAIA is applicable if $b_{i,j}=0$ (Figure~\ref{fig:energy_LZSM}(b)). 

Because we truncate the dimension of the boson Hilbert space, the transition probabilities can be adequately approximated within a finite number of crossings. In the following, we compare the results of GAIA with the exact results (numerical calculations) when $20$ crossings occur (Figure~\ref{fig:prob_LZSM_02_1}). The initial state is set to be the ground state of the initial Hamiltonian $\ket{1}$.

First, we consider the case of $\sqrt{\eta/v} a=0.2$. For example, in the case of $\gamma=0.02\sqrt{v}$, we find that the results agree well with the numerical results within the range of $20$ crossings (Figure~\ref{fig:prob_LZSM_02_1}(a)). On the other hand, in the case of $\gamma=0.1\sqrt{v}$, the numerical results cannot be approximated well because of the large off-diagonal elements  (Figure~\ref{fig:prob_LZSM_02_1}(b)). Next, we consider the case of $\sqrt{\eta/v}a=1.0$. In this case, we find that the results agree well with the numerical results in the case not only of $\gamma=0.02\sqrt{v}$ but also of $\gamma=0.1\sqrt{v}$ (Figure~\ref{fig:prob_LZSM_02_1}(c,d)). This can be explained by the fact that the time intervals of the anticrossing become larger, which means that the conditions for GAIA are satisfied. We note that since the dimension of the boson Hilbert space $N$ is finitely truncated in the numerical calculation, the dynamics, especially after the time when the probability amplitude of the $(2N-1)$th excited state is large, is different from the dynamics under the actual Hamiltonian~\eqref{eq:H_spinboson}.

\begin{figure}[t]
    \centering
    \includegraphics[width=85mm]{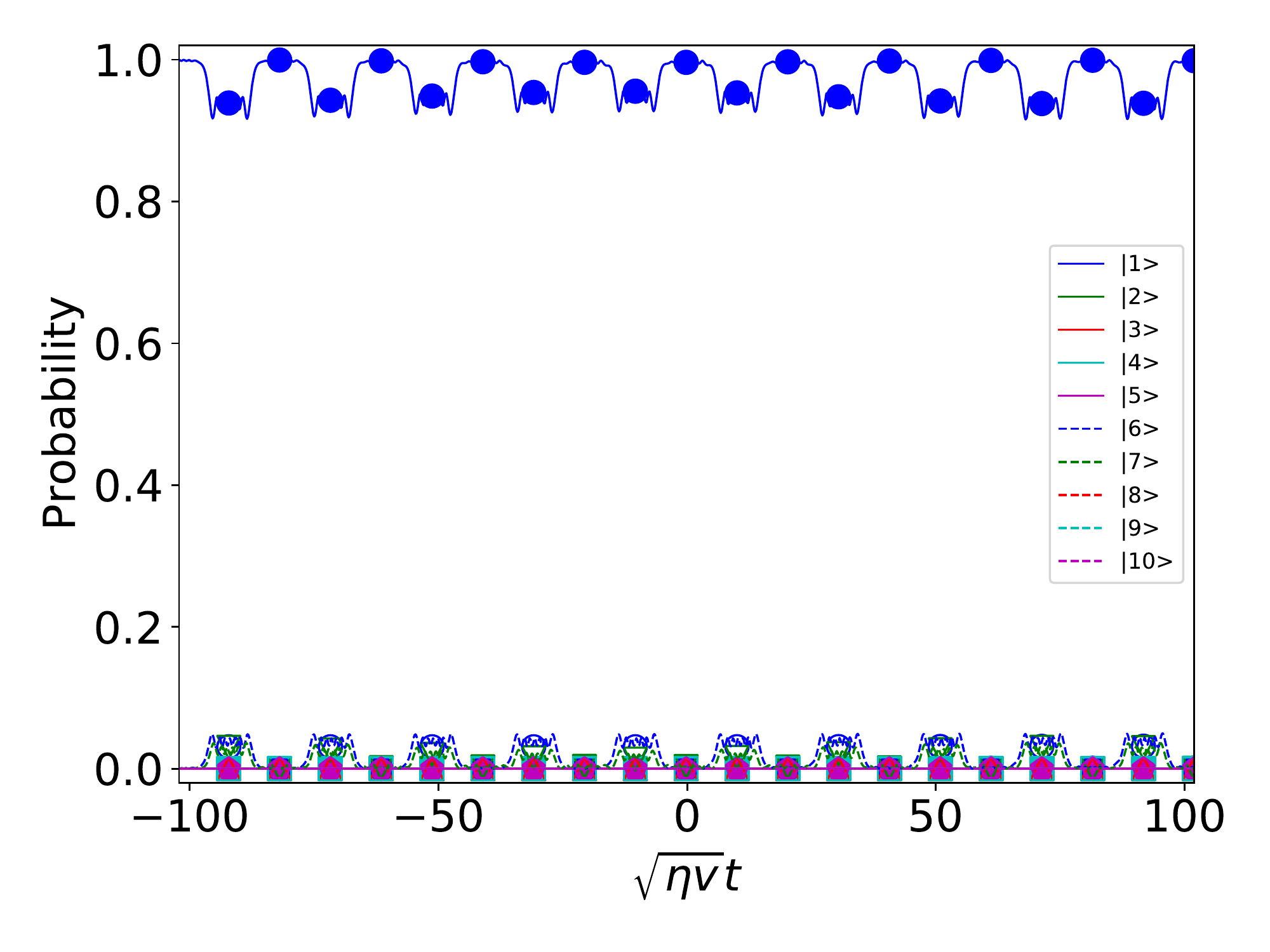}
    \caption{Time evolution when the conditions for destructive interference are satisfied. The parameters are $\Delta=\gamma=0.1\sqrt{v}$, $\sqrt{\eta/v}a=0.2\sqrt{v}$, and $\eta=10.57 v$, which are satisfied with \eqref{eq:destructive_condition}.}
    \label{fig:prob_LZSM_destructive}
\end{figure}

The interference plays an important role in the periodical model~\cite{Du2010-sf,Bonifacio2020-lr,Lidal2020-go,Neilinger2016-jw,Wang2021-fy}. Although the conditions for interference in the two-level system have been studied, the conditions for interference in the multilevel system have not been studied straightforwardly so far. By using the S-matrix here obtained, the conditions for interference can be derived. For example, if the initial state is the ground state, we can derive the condition of destructive interference for the Hamiltonian~\eqref{eq:H_spinboson} which means $S_{11}=1$ after $2$ crossings occur. 
When $2$ crossings occur, the transition amplitude between $\ket{1}$ becomes
\begin{align}
    S_{11}=&(1-p_{1,N+1}) e^{i\left(\theta_{1,N+1, 1}+\theta_{1,N+1, 2}\right)}\\
    &+p_{1,N+1} (1-p_{1,N+2}) e^{i\left(\theta_{1,N+2, 1}+\theta_{1,N+2, 2}\right)}\\
    &+p_{1,N+1} p_{1,N+2}.\label{eq:destructive_condition}
\end{align}
If the conditions $\theta_{1,N+1, 1}+\theta_{1,N+1, 2}=2\mathbb{Z}\pi$ and $\theta_{1,N+2, 1}+\theta_{1,N+2, 2}=2\mathbb{Z}\pi$ hold, $S_{11}=1$ and a destructive interference occurs (Figure~\ref{fig:prob_LZSM_destructive}). We note that we have truncated the dimension of the boson, but this does not affect the conditions for destructive interference.

\section{Conclusion\label{sec:conclusion}}
In this study, we propose a GAIA by improving the S-matrix obtained in previous studies~\cite{Aoki2002-zw,shimada2020numerical} and applied the results to the LZ grid model and the LZSM interference model. The division of the S-matrix into unitary matrices, not only makes numerical calculations easier, but also allows for a physical interpretation. In the LZ grid model, we have investigated the relation between the S-matrix obtained by GAIA and the conventional AIA. In addition, we have analyzed the LZSM interference model, for which the previous methods~\cite{Aoki2002-zw,shimada2020numerical} have difficulties to analyze, and a new condition for interference is obtained.

The present results can be applied to other multilevel LZ model, for which integrability conditions have been studied~\cite{Sinitsyn2017-xn,Sinitsyn2018-pa,Chernyak2018-uz,Chernyak2020-kx,Chernyak2020-pm,Chernyak2021-ph}. We expect that the present method can be used to obtain the necessary condition for integrability. In the integrable multilevel LZ model, crossing three or more levels at a single point were sometimes considered. Since GAIA cannot calculate the S-matrix of such a model, it is necessary to extend the present method.
We note that GAIA can describe well the dynamics when non-adiabatic transitions generate interference. We expect that GAIA can be applied to quantum information processing in such a setting~\cite{Hanggi2007-fo,Deng2016-yg,Munoz-Bauza2019-po}.

\section*{Acknowledgement}
This work was supported by JSPS KAKENHI Grant Number JP21J12952. T.S. was supported by Waseda Research Institute for Science and Engineering, Grant-in-Aid for Young Scientists (Early Bird) and Top Global University Project, Waseda University.
H.N. is supported in part by the Institute for Advanced Theoretical and Experimental
Physics, Waseda University and by Waseda University Grant for Special Research Projects (Project number 2021C-196).

\appendix
\section{Unitary Decomposition\label{sec:app1}}

To express the S matrix as a product of unitary matrices, we define the matrix $\tilde U_k$ as follows:
\begin{align}
    \tilde M_k \tilde N_{k}^{(-)}=&\tilde N_{k-1}^{(-)} \tilde U_k,\quad (\tilde N_{2N-1}^{(-)}=\tilde N^{(-)}) \label{eq:unitary},
\end{align}
where
\begin{widetext}
\begin{align}
    \tilde N_{k}^{(-)}=\begin{dcases}
    \operatorname{diag}\Biggl(\overbrace{\prod_{l=N+1}^{k+1} p_{1 l}^{-1 / 2}, \cdots,\prod_{l=N+1}^{2N} p_{2N-k-1 ,l}^{-1 / 2},}^{2N-k-1} \overbrace{ \prod_{l=N+1}^{2 N} p_{2N-k, l}^{-1 / 2},\cdots, \prod_{l=N+1}^{2 N} p_{N l}^{-1 / 2}}^{k-N+1},\\
    \quad \quad \quad \vec I_{k-N+1}, \overbrace{\prod_{l=1}^{1} p_{l, k+2}^{-1 / 2}, \cdots, \prod_{l=1}^{2N-k-1} p_{l, 2 N}^{-1 / 2}}^{2N-k-1}\Biggr)&(k\geq N)\\
    \operatorname{diag}\Biggl(\vec I_{N-k}, \overbrace{\prod_{l=N+1}^{N+1} p_{N-k+1, l}^{-1 / 2}, \cdots, \prod_{l=N+1}^{N+k} p_{N, l}^{-1 / 2},}^{k} \\
    \quad \quad \quad \overbrace{\prod_{l=1}^{N-k} p_{l, N+1}^{-1 / 2}, \cdots,  \prod_{l=1}^{N-1} p_{l, N+k}^{-1 / 2},}^{k} \overbrace{  \prod_{l=1}^{N} p_{l, N+k+1}^{-1 / 2},\cdots \prod_{l=1}^{N} p_{l, 2 N}^{-1 / 2}}^{N-k}\Biggr)&(k\leq  N)
    \end{dcases}\label{eq:nk_minus},
\end{align}
and we also define
\begin{align*}
    \vec I_n=(\overbrace{1,\cdots,1}^{n}).
\end{align*}

The matrix $M_{i,j}$ can be transformed using any $X_{i,j},Y_{i,j}\in\mathbb{C}$ as follows:
\begin{align*}
    M_{i,j}=&\left(\begin{array}{ccccc}
    I_{i-1} & & & & \\
     & X^{1/2}_{i,j} & \cdots & 0 & \\
     & \vdots & I_{j-i-1} & \vdots & \\
     & 0 & \cdots & X^{-1/2}_{i,j}Y_{i,j}^{-1} & \\
     & & & & I_{2 N-j}
    \end{array}\right)\\
    &\times\left(\begin{array}{ccccc}
    I_{i-1} & & & & \\
     & p_{i ,j}X_{i,j}^{-1} & \cdots & -\alpha_{i ,j}^{+}Y^{-1}_{i,j} & \\
     & \vdots & I_{j-i-1} & \vdots & \\
     & -\alpha_{i ,j}^{-}Y_{i,j} & \cdots & X_{i,j} & \\
     & & & & I_{2 N-j}
    \end{array}\right)
     \left(\begin{array}{ccccc}
    I_{i-1} & & & & \\
     & X_{i,j}^{1/2} & \cdots & 0 & \\
     & \vdots & I_{j-i-1} & \vdots & \\
     & 0 & \cdots & X^{-1/2}_{i,j}Y_{i,j} & \\
     & & & & I_{2 N-j}
    \end{array}\right)\\
    =:& M_{i,j}^{(-)} \tilde U_{i,j}  M_{i,j}^{(+)}.
\end{align*}
Hereafter, we define $X_{i,j}=p_{i,j}^{1/2},\ Y_{i,j}=\prod_{l=1}^{i-1}p_{lj}^{1/2}\prod_{l=N+1}^{j-1}p_{il}^{-1/2}$. The left hand side of \eqref{eq:unitary} can be transformed like 
\begin{align*}
    \tilde M_{k}\tilde N_{k}^{(-)}=&\prod_{\substack{j-i=k \\ (1\leq i\leq N<N+1\leq j\leq 2N)}}M_{i,j}\tilde N_{k}^{(-)}
    =\begin{dcases}
    \prod_{1\leq i\leq 2N-k}M_{i,i+k}\tilde N_{k}^{(-)}&(k\geq N)\\
    \prod_{N-k+1\leq i\leq N}M_{i,i+k}\tilde N_{k}^{(-)}&(k\leq N) \label{eq:MNminus}
    \end{dcases}.
\end{align*}
First, we consider the case of $k\geq N$. We decompose $\tilde N_{k}^{(-)}$ into two parts:
\begin{align*}
    \tilde N_{k}^{(-)}=&\left(\prod_{1\leq i\leq 2N-k} N_{i,i+k}^{(-)}\right)\tilde N_k^{(-),\perp},\\
    N_{i,i+k}^{(-)}:=&\operatorname{diag}\Biggl(\vec I_{i-1}, \prod_{l=N+1}^{i+k} p_{i, l}^{-1 / 2},\vec I_{k-1}, \prod_{l=1}^{i-1} p_{l, i+k}^{-1 / 2},\vec I_{2N-i-k}\Biggr),\\
    \tilde N_k^{(-),\perp}:=&\operatorname{diag}\Biggl(\vec I_{2N-k},\overbrace{\prod_{l=N+1}^{2 N} p_{2N-k+1, l}^{-1 / 2}, \cdots, \prod_{l=N+1}^{2 N} p_{N, l}^{-1 / 2}}^{k-N},\vec I_{N}\Biggr).
\end{align*}
Then, we can compute like this:
\begin{align*}
     M_{i,i+k}^{(+)} N_{i,i+k}^{(-)}
    =&\operatorname{diag}\left(\vec I_{i-1}, X_{i,i+k}^{1/2}\prod_{l=N+1}^{i+k} p_{i, l}^{-1 / 2},\vec I_{k-1},X_{i,i+k}^{1/2}\prod_{l=N+1}^{i+k} p_{i, l}^{-1 / 2},\vec I_{2N-i-k}\right),\\
    \therefore \ M_{i,i+k}^{(-)}\tilde U_{i,i+k}  M_{i,i+k}^{(+)} N_{i,i+k}^{(-)}=&M_{i,i+k}^{(-)} M_{i,i+k}^{(+)} N_{i,i+k}^{(-)}\tilde U_{i,i+k} \\
    =&\operatorname{diag}\left(\vec I_{i-1}, \prod_{l=N+1}^{i+k-1} p_{i, l}^{-1 / 2},\vec I_{k-1}, \prod_{l=1}^{i} p_{l, i+k}^{-1 / 2},\vec I_{2N-i-k}\right)\tilde U_{i,i+k}\\
    =:& N'_{i,i+k}\tilde U_{i,i+k}.
\end{align*}
The left hand side of \eqref{eq:unitary} can be transformed like 
\begin{align*}
    \tilde M_{k}\tilde N_{k}^{(-)}=&\prod_{1\leq i\leq 2N-k}( M_{i,i+k}^{(-)} \tilde U_{i,i+k}  M_{i,i+k}^{(+)} N_{i,i+k}^{(-)})N_k^{(-),\perp}\\
    =&\tilde N_k^{(-),\perp}\prod_{1\leq i\leq 2N-k} N'_{i,i+k} \prod_{1\leq i\leq 2N-k}\tilde U_{i,i+k}.
\end{align*}
We used the relation $[N'_{i,i+k},\tilde U_{j,j+k}]=0\ (i\neq j)$ in the above transformation. Finally, we calculate
\begin{align*}
    \tilde N_k^{(-),\perp}\prod_{1\leq i\leq 2N-k} N'_{i,i+k}=&\operatorname{diag}\Biggl(\vec I_{2N-k},\overbrace{\prod_{l=N+1}^{2 N} p_{2N-k+1, l}^{-1 / 2}, \cdots, \prod_{l=N+1}^{2 N} p_{N, l}^{-1 / 2}}^{k-N},\vec I_{N}\Biggr)\\
    &\times \prod_{1\leq i\leq 2N-k}\operatorname{diag}\left(\vec I_{i-1}, \prod_{l=N+1}^{i+k-1} p_{i, l}^{-1 / 2},\vec I_{k-1}, \prod_{l=1}^{i} p_{l, i+k}^{-1 / 2},\vec I_{2N-i-k}\right)\\
    =&\operatorname{diag}\Biggl(\overbrace{\prod_{l=N+1}^{k} p_{1, l}^{-1 / 2}, \cdots, \prod_{l=N+1}^{2N-1} p_{2N-k, l}^{-1 / 2}}^{2N-k},\overbrace{\prod_{k=N+1}^{2 N} p_{2N-k+1, k}^{-1 / 2}, \cdots, \prod_{k=N+1}^{2 N} p_{N, k}^{-1 / 2}}^{k-N},\\
    &\quad\quad\quad\vec I_{k-N},\overbrace{\prod_{l=1}^{1} p_{l, k+1}^{-1 / 2},\cdots,\prod_{l=1}^{2N-k} p_{l, 2N}^{-1 / 2}}^{2N-k}\Biggr)\\
    =&\tilde N_{k-1}^{(-)}.
\end{align*}
\end{widetext}
Then, we proved \eqref{eq:unitary} in the case of $k\geq N$.

Next, we consider the case of $k\leq N$. We decompose $\tilde N_{k}^{(-)}$ into two parts:
\begin{align*}
    \tilde N_{k}^{(-)}=&\left(\prod_{N-k+1\leq i\leq N} N_{i,i+k}^{(-)}\right)\tilde N_k^{(-),\perp}
\end{align*}
\begin{align*}
    N_{i,i+k}^{(-)}=&\operatorname{diag}\Biggl(\vec I_{i-1}, \prod_{l=N+1}^{i+k} p_{i, l}^{-1 / 2}, \vec I_{k-1}, \prod_{l=1}^{i-1} p_{l, i+k}^{-1 / 2},\vec I_{2N-i-k}\Biggr)\\
    \tilde N_k^{(-),\perp}=&\operatorname{diag}\Biggl(\vec I_{N+k}, \overbrace{  \prod_{l=1}^{N} p_{l, N+k+1}^{-1 / 2},\cdots \prod_{l=1}^{N} p_{l, 2 N}^{-1 / 2}}^{N-k}\Biggr)
\end{align*}
\begin{widetext}
Then, we can compute
\begin{align*}
     M_{i,i+k}^{(+)} N_{i,i+k}^{(-)}
    =&\operatorname{diag}\left(\vec I_{i-1}, X_{i,i+k}^{1/2}\prod_{l=N+1}^{i+k} p_{i, l}^{-1 / 2},\vec I_{k-1},X_{i,i+k}^{1/2}\prod_{l=N+1}^{i+k}p_{il}^{-1/2}  ,\vec I_{2N-i-k}\right),\\
    \therefore \ M_{i,i+k}^{(-)}\tilde U_{i,i+k}  M_{i,i+k}^{(+)} N_{i,i+k}^{(-)}=&M_{i,i+k}^{(-)} M_{i,i+k}^{(+)} N_{i,i+k}^{(-)}\tilde U_{i,i+k} \\
    =&\operatorname{diag}\left(\vec I_{i-1}, \prod_{l=N+1}^{i+k-1} p_{i, l}^{-1 / 2},\vec I_{k-1}, \prod_{l=1}^{i} p_{l, i+k}^{-1 / 2},\vec I_{2N-i-k}\right)\tilde U_{i,i+k}\\
    =:& N'_{i,i+k}\tilde U_{i,i+k}.
\end{align*}
The left hand side of \eqref{eq:unitary} can be transformed like 
\begin{align*}
    \tilde M_{k}\tilde N_{k}^{(-)}=&\prod_{N-k+1\leq i\leq N}( M_{i,i+k}^{(-)} \tilde U_{i,i+k}  M_{i,i+k}^{(+)} N_{i,i+k}^{(-)})N_k^{(-),\perp}\\
    =&\tilde N_k^{(-),\perp}\prod_{N-k+1\leq i\leq N}(  N'_{i,i+k} \tilde U_{i,i+k}).
\end{align*}
Finally, we calculate
\begin{align*}
    \tilde N_k^{(-),\perp}\prod_{N-k+1\leq i\leq N} N'_{i,i+k}=&\operatorname{diag}\Biggl(\vec I_{N+k}, \overbrace{  \prod_{l=1}^{N} p_{l, N+k+1}^{-1 / 2},\cdots \prod_{l=1}^{N} p_{l, 2 N}^{-1 / 2}}^{N-k}\Biggr)\\
    &\times \prod_{N-k+1\leq i\leq N}\operatorname{diag}\left(\vec I_{i-1}, \prod_{l=N+1}^{i+k-1} p_{i, l}^{-1 / 2},\vec I_{k-1}, \prod_{l=1}^{i} p_{l, i+k}^{-1 / 2},\vec I_{2N-i-k}\right)\\
    =&\operatorname{diag}\Biggl(\vec I_{N-k},\overbrace{1,\prod_{l=N+1}^{N+1} p_{N-k+2, l}^{-1 / 2} \cdots, \prod_{l=N+1}^{N+k-1} p_{N, l}^{-1 / 2}}^{k},\\
    &\quad\quad\quad\overbrace{ \prod_{l=1}^{N-k+1} p_{l, N+1}^{-1 / 2},\cdots, \prod_{l=1}^{N} p_{l, N+k}^{-1 / 2}}^{k},\overbrace{  \prod_{l=1}^{N} p_{l, N+k+1}^{-1 / 2},\cdots \prod_{l=1}^{N} p_{l, 2 N}^{-1 / 2}}^{N-k}\Biggr)\\
    =&\tilde N_{k-1}^{(-)}.
\end{align*}
\end{widetext}
Then, we proved \eqref{eq:unitary} in the case of $k\leq N$. 

Using the relation \eqref{eq:unitary}, \eqref{eq:prev_S} can be transformed as
\begin{align*}
    S=&\left(\tilde N^{(+)}\right)^{-1} \tilde M_{1}\cdots\tilde M_{2N-2}\tilde M_{2N-1} \tilde N^{(-)}_{2N-1}\\
    =&\left(\tilde N^{(+)}\right)^{-1} \tilde N^{(-)}_0\tilde U_{1}\cdots\tilde U_{2N-2}\tilde U_{2N-1}.
\end{align*}
From \eqref{eq:nk_minus} and \eqref{eq:n_plus}, $\tilde N^{(+)}=\tilde N_0^{(-)}$ holds. Then, we proved \eqref{eq:S-matrix}.

\bibliographystyle{apsrev4-1}
\bibliography{ref}

\end{document}